\documentclass[lettersize,journal]{IEEEtran}
\usepackage[utf8]{inputenc}
\usepackage{amsmath,amsfonts,amssymb}
\usepackage{algorithm}
\usepackage{algorithmic}

\usepackage{bm}
\usepackage{array}
\usepackage[caption=false,font=footnotesize,labelfont=rm,textfont=rm]{subfig}
\usepackage{textcomp}
\usepackage{stfloats}
\newtheorem{theorem}{Theorem}

\usepackage{url}
\usepackage{verbatim}
\usepackage{graphicx}
\usepackage{lipsum}
\usepackage{cite}
\usepackage{hyperref}
\hypersetup{hypertex=true,
	colorlinks=true,
	linkcolor=black,
	anchorcolor=black,
	citecolor=blue}
\allowdisplaybreaks[1]
\makeatletter
\renewcommand{\maketag@@@}[1]{\hbox{\m@th\normalsize\normalfont#1}}%
\makeatother
\begin{document}

\title{Movable-Antenna Array Enhanced Multi-Target Sensing: CRB Characterization and Optimization}

\author{{Haobin Mao},~\IEEEmembership{Graduate Student Member,~IEEE,}
	{Lipeng Zhu},~\IEEEmembership{Member,~IEEE,}
	{Wenyan Ma},~\IEEEmembership{Graduate Student Member,~IEEE,}
	{Zhenyu Xiao},~\IEEEmembership{Senior Member,~IEEE,}
	{Xiang-Gen Xia},~\IEEEmembership{Fellow,~IEEE,}
	and {Rui Zhang},~\IEEEmembership{Fellow,~IEEE}
	
	\thanks{H. Mao and Z. Xiao are with the School of Electronic and Information Engineering and the State Key Laboratory of CNS/ATM, Beihang University, Beijing 100191, China (e-mail: maohaobin@buaa.edu.cn, xiaozy@buaa.edu.cn). \textit{(Corresponding authors: Lipeng Zhu and Zhenyu Xiao.)}} 
	\thanks{L. Zhu, W. Ma, and R. Zhang are with the Department of Electrical and Computer Engineering, National University of Singapore, Singapore 117583 (e-mail: zhulp@nus.edu.sg, wenyan@u.nus.edu, elezhang@nus.edu.sg).}
	\thanks{X.-G. Xia is with the Department of Electrical and Computer Engineering, University of Delaware, Newark, DE 19716, USA (e-mail: xxia@ee.udel.edu).}}


\maketitle
\begin{abstract} 
Movable antennas (MAs) have emerged as a promising technology to improve wireless communication and sensing performance towards sixth-generation (6G) networks through flexible antenna movement. In this paper, we propose a novel wireless sensing system based on MA arrays to enhance multi-target spatial angle estimation performance. We begin by characterizing the Cram\'{e}r-Rao bound (CRB) matrix for multi-target angle of arrival (AoA) estimation as a function of the antenna's positions in MA arrays, thereby establishing a theoretical foundation for antenna position optimization. Then, aiming at improving the sensing coverage performance, we formulate an optimization problem to minimize the expectation of the trace of the CRB matrix over random target angles subject to a given distribution by optimizing the antennas' positions. To tackle the formulated challenging optimization problem, the Monte Carlo method is employed to approximate the intractable objective function, and a swarm-based gradient descent algorithm is subsequently proposed to address the approximated problem. In addition, a lower-bound on the sum of CRBs for multi-target AoA estimation is derived. Numerical results demonstrate that the proposed MA-based design achieves superior sensing performance compared to conventional systems using fixed-position antenna (FPA) arrays and single-target-oriented MA arrays, in terms of decreasing both CRB and the actual AoA estimation mean square error (MSE). Fundamentally, the designed MA array geometry exhibits low correlation and high effective power of sensitivity vectors for multi-target sensing in the angular domain, leading to significant CRB performance improvement. The resultant low correlation of steering vectors over multiple targets' directions further helps mitigate angle estimation ambiguity and thus enhances MSE performance.
\end{abstract}

\vspace{-0.8em}
\section{Introduction}
\IEEEPARstart{T}{he} envisioned sixth-generation (6G) wireless networks are expected to transcend traditional communication paradigms by integrating sensing as a native functionality, enabling diverse location-aware applications across domains such as autonomous driving, smart cities, and low-altitude economy \cite{jiang20216g}. In this context, high-precision environmental sensing, encompassing accurate angle/location estimation and extraction of physical information from targets, are anticipated to become essential features of next-generation wireless infrastructure \cite{zhu20256g}. Toward this goal, large-scale antenna arrays have been widely explored to enhance wireless sensing performance, albeit with associated increases in hardware cost and power consumption. To mitigate these practical deployment constraints, sparse antenna arrays have been introduced with larger inter-antenna spacing to achieve fine spatial resolution with a reduced number of antenna elements \cite{robert2011sparse,li2025sparse}. However, the adoption of fixed-position antennas (FPAs) impedes the full exploitation of the degrees of freedom (DoFs) in the spatial domain, thereby limiting the adaptivity for varying sensing requirements in wireless networks.

To address the above limitations of FPA arrays for wireless sensing, movable antennas (MAs) have recently emerged as a prospective technology to enhance wireless sensing performance via flexible antenna position adjustment  \cite{zhu2025tutorial,ma2024sensing,ma2025trajectory}, which are also known as fluid antennas \cite{zhu2024historical} or flexible antennas \cite{zheng2024flexible}. Compared to conventional fixed-position antenna (FPA) arrays, MA arrays unlock additional DoFs by enabling adaptive antenna positioning within a spatial region. This facilitates substantial sensing performance gains with the same or even reduced number of antennas compared to FPA arrays. Specifically, the effective array aperture can be enlarged through dynamic antenna position adjustment, which directly enhances angle estimation resolution. In addition, the geometry of the MA array can be optimized to suppress grating and side lobes in undesired directions, thereby reducing inter-target coupling and mitigating angle estimation ambiguity \cite{ma2024sensing}. Moreover, the real-time trajectory control of a limited number of MAs allows the formation of virtual antenna arrays with reconfigurable apertures, offering on-demand support for evolving sensing tasks \cite{ma2025trajectory}.

In addition to wireless sensing, MAs have also demonstrated superior performance in wireless communication systems. The inception of MA-enabled wireless communication dates back to 2009 \cite{zhao2009ma}, where significant spatial diversity gain was obtained via antenna movement within a predefined geographical region. In recent years, especially after 2023, a substantial number of studies have validated the performance advantages of MAs over their FPA counterparts across diverse wireless applications ranging from terrestrial communications to unmanned aerial vehicle (UAV) and satellite communications, with the fundamental gains originating from spatial diversity \cite{zhu2024movable,zhu2024modeling,zhou2024noma,mei2024graph}, flexible beamforming \cite{hu2024secure,liu2024uav,zhu2024nearfield}, and multiplexing enhancement \cite{zhu2024multiuser,ding2025movable,xiao2024multiuser,xiao2025noma} via antenna position reconfiguration. As an extension, six-dimensional MA (6DMA) systems incorporating joint antenna position and rotation adjustment have been proposed to further enhance the performance of wireless networks \cite{shao20246dma,shao20256dma,shao2025discre}. As a simplified architecture with antenna rotation only, the rotatable antenna technology has also shown considerable superiority to balance between hardware complexity and system performance \cite{zheng2025modeling,zheng2025rotatable}. In parallel, research efforts have also been dedicated to acquiring accurate channel state information (CSI) for MA systems in \cite{zhang2024channel,xiao2024channel,shao2025distri} to enable antenna position/rotation optimization. To further increase the spatial DoFs, extremely large-scale MA (XL-MA) architectures have been proposed by enabling antenna/subarray movement within large regions (e.g., on the order of several to tens of meters) \cite{fu2025extremely}, which share a similar idea to the pinching antennas to proactively create line-of-sight (LoS) channels between transceivers and reduce their path loss \cite{ding2024flexible,liu2025pinching,liu2025tutorial,wang2025modeling,ouyang2025capacity}.

More recently, preliminary efforts have been directed toward deploying MAs in wireless sensing systems, aiming to capitalize on their spatial DoFs for enhanced sensing performance. For instance, the authors in \cite{ma2024sensing} characterized the Cram\'{e}r-Rao bound (CRB) for single target's angle of arrival (AoA) estimation, which was then minimized via optimizing MAs' positions. The authors in \cite{ma2025trajectory} extended their work by exploiting the additional time DoF provided by optimizing the MA trajectory to synthesize a large continuous virtual array for improving sensing performance. The authors in \cite{li2025sensing} optimized the antenna array geometry at both the transmitter and the receiver as well as the signal covariance to minimize the weighted sum of CRBs. Furthermore, the authors in \cite{wang2025antenna} employed MAs for improving both the angle and range estimation performance in near-field scenarios via antenna position adjustment.

Despite these advancements, the above-mentioned works have primarily focused on single-target sensing tasks. Consequently, the solutions proposed in \cite{ma2024sensing,ma2025trajectory,li2025sensing,wang2025antenna} may underperform in multi-target sensing systems. In multi-target scenarios, the estimation for the angle of one target is critically dependent on the angles of other targets, and the inherent coupling effect introduces distinct challenges for MA-aided sensing system design. In this context, the authors in \cite{ding2025masensing} maximized the conditional sensing mutual information for multi-target sensing via joint optimization of receive combining, sensing signal covariance matrices, transmit beamforming, and MAs' positions. The authors in \cite{wu2025movable} jointly optimized the MAs' positions, transmit beamforming, and the phase shifts of reconfigurable intelligent surface (RIS) to maximize the beampattern gain in the direction of multiple targets. Although these prior works have demonstrated multi-target sensing performance improvements through MA positioning, the fundamental relationship between MAs’ positions and sensing performance limit for multi-target angle estimation has not been revealed yet.

To bridge the above gap, in this paper, we propose to employ an MA array to enhance multi-target sensing performance through array geometry reconfiguration by characterizing the CRB matrix as a function of antennas' positions. The main contributions of this paper are summarized as follows:

\begin{itemize}
	\item[1)] We present a novel MA array enhanced multi-target sensing system, where a base station (BS) equipped with an MA array to estimate the spatial AoAs from multiple targets based on the received source signals. First, we derive the CRB matrix for multi-target AoA estimation and characterize it as a function of the antenna positions within the MA array, thereby establishing a theoretical foundation for antenna position optimization. Then, aiming at improving the sensing coverage performance, we formulate an optimization problem to minimize the expected sum of the CRBs for multi-target AoA estimation (i.e., the trace of the CRB matrix) over random target distributions by optimizing the antennas' positions.
	
	\item[2)] To address the formulated highly non-convex optimization problem, the Monte Carlo method is utilized to approximate the intractable objective function, followed by the development of a swarm-based gradient descent algorithm to solve the approximated problem. Furthermore, a lower-bound on the sum of CRBs for multi-target AoA estimation is derived, where we reveal the conditions to achieve this lower-bound.
	
	\item[3)] Numerical results validate that the proposed MA-based design achieves superior sensing performance over conventional systems with FPA arrays and single-target-oriented MA arrays, in terms of decreasing both the CRB and the actual AoA estimation mean square error (MSE). This is because the optimized MA array geometry exhibits low correlation and high effective power of sensitivity vectors for multi-target sensing in the angular domain, contributing to considerable improvements in the CRB performance. The resultant low correlation of steering vectors over multiple targets' directions further helps alleviate angle estimation ambiguity and thus leads to significant gains in the MSE performance. These results highlight the significance of our design in enabling high-precision multi-target angle estimation.
	
\end{itemize}

The rest of this paper is organized as follows. In Section \ref{section:s2}, we present the system model,
charecterize the CRB matrix of multi-target AoA estimation MSE, and formulate the optimization problem. In Section \ref{section:s3}, we propose a swarm-based gradient descent algorithm to solve the formulated problem, where the computational complexity and the performance lower-bound are also analyzed. In Section \ref{section:s4}, we provide numerical results to evaluate the performance of our proposed design. Finally, we conclude this paper in Section \ref{section:s5}.

\textit{Notation}: $a, \bf a$, and $\bf A$ denote a scalar, a vector, and a matrix, respectively. $(\cdot)^{\sf T}$ and $(\cdot)^{\sf H}$ denote transpose and conjugate transpose, respectively. ${\mathbb{R}}^{M \times N}$ and ${\mathbb{C}}^{M \times N}$ denote the sets of real and complex matrices/vectors with dimension $M \times N$, respectively. $\Re\{ \cdot \}$ and $\Im (\cdot)$ indicate the real part and the imaginary part of a complex scalar/vector/matrix, respectively. $\Vert \cdot \Vert _2$ denotes the $l_2$-norm of a vector. ${{\bf I}}_N$ denotes the identity matrix with the dimension $N\times N$. ${\mathbf{e}}_n$ denotes the $N$-dimensional vector with the $n$-th entry being $1$ and all other entries being zero. ${\bf 1}_N$ indicates the matrix with the dimension $N\times N$ with all the entries being 1.  ${{\bf 0}}_{N \times 1}$ and ${{\bf 0}}_{N}$ denote the vector with the dimension $N\times 1$ and the matrix with the dimension $N\times N$ with all the entries being zero, respectively. $\mathcal{CN}\left({\bf{0}}_{N\times 1},{\sigma}^2{{\bf I}}_N\right)$ denotes the circularly symmetric complex Gaussian (CSCG) distribution with zero mean and covariance matrix ${\sigma}^2{\bf I}_N$. $\mathbb{E} \{ \cdot \}$ represents the expectation operation of a random variable. $\otimes$ and $\odot$ denote the Kronecker product and the Hadamard product, respectively. $\nabla_{{\bf x}} f ( {\bf x}_0)$ denotes the gradient vector of function $f ( \bf x)$ at the local point ${\bf x}_0$. $\partial(\cdot)$ represents the partial derivative of a function. $\left[{\bf a}\right]_i$ and $\left[{\bf A}\right]_{i,j}$ denote the $i$-th entry of vector ${\bf a}$ and the entry in the $i$-th row and the $j$-th column of matrix $\bf A$, respectively. ${\rm tr}({\bf A})$ denotes the trace of matrix $\bf A$.
\vspace{-0.5em}
\section{System Model and Problem Formulation} \label{section:s2}
\subsection{System Model}
\begin{figure}[t]
	\begin{center}
		\includegraphics[width= 7 cm]{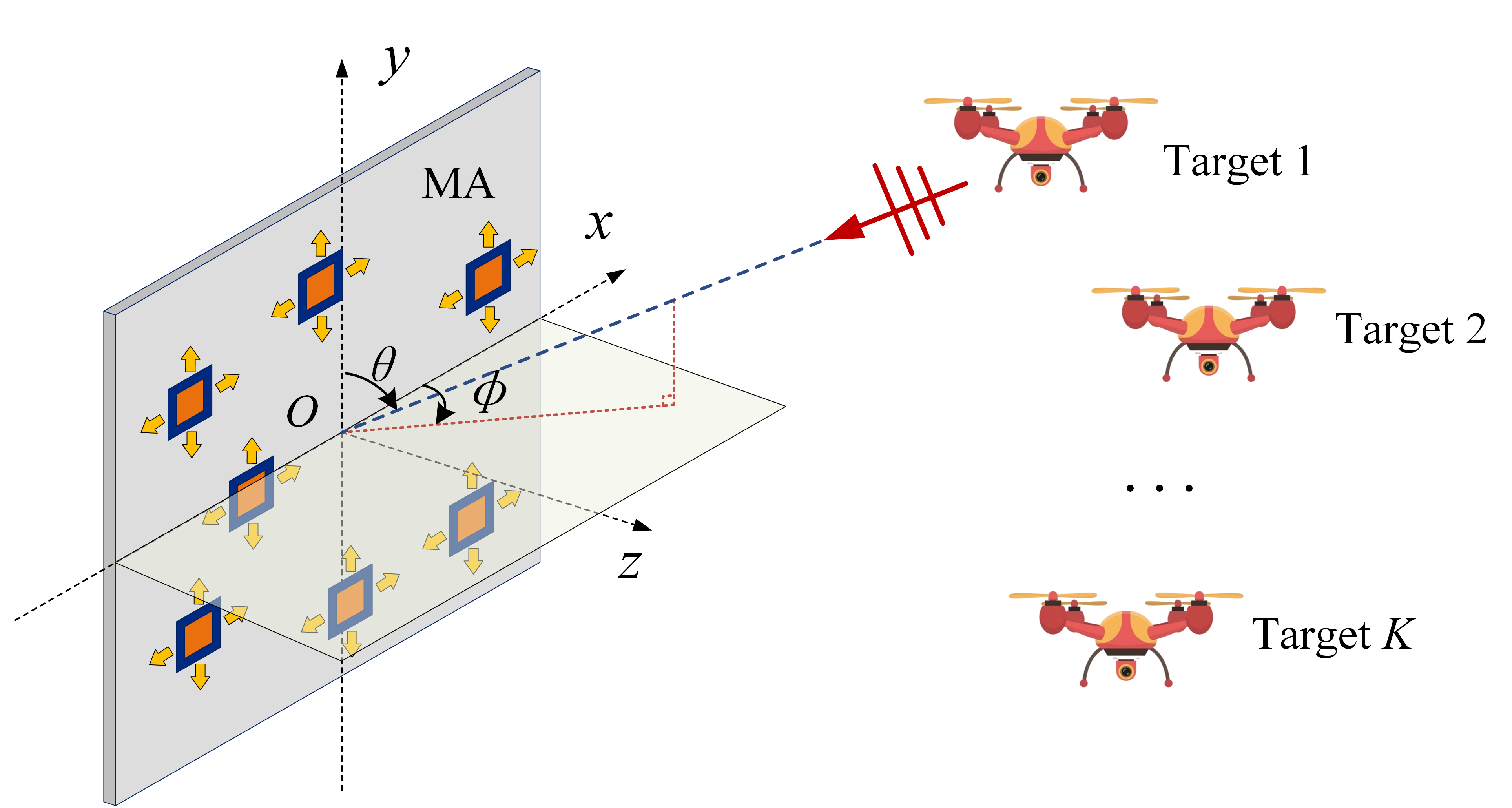}
		\caption{Illustration of the considered MA array enhanced multi-target sensing system.}
		\label{fig:Scenario}
	\end{center}
\vspace{-2em}
\end{figure}

As shown in Fig. \ref{fig:Scenario}, we consider a passive wireless sensing system with $N$ MAs confined within a two-dimensional (2D) region $\mathcal C$ at the BS to estimate $K$ targets' AoAs with respect to (w.r.t.) $x$ and $y$ axes\footnote{It is worth noting that the proposed solution in this paper for the passive wireless sensing system is also applicable to active wireless sensing systems where the transmitter emits omnidirectional probing signals.}, where the antenna moving region is assumed to be a square with the size of $A \times A$. Assume that the number of antennas is larger than the number of targets, i.e., $N>K$. To describe the positions of MAs, we establish a Cartesian coordinate system (CCS) with the reference point of the antenna array located at the origin $O$. The position of the $n$-th antenna can then be denoted as ${{\bf{q}}_n} = {[{x_n},{y_n}]^{\mathsf T}}$ for $1 \le n \le N$. For brevity, we denote the collection of $N$ MAs' positions as $\tilde{{\bf q}} =\{ {{\bf{q}}_n},1\le n \le N\}$.

To perform AoA estimation, the receiver consecutively receives the uncorrelated narrowband signals from multiple targets over $T$ snapshots \cite{esfandiari2025doa}. We assume that the channels from the targets to the receiver are dominated by the LoS components and remain static over $T$ snapshots. Given that multiple targets are located in the far-field region of the MA array, the steering vector from the MA array to the $k$-th target is given by
\begin{equation}\small
{\bf{a}}\left( \tilde{{\bf q}},{{\bf{r}}_k} \right) = {\left[ {{{\rm{e}}^{{\rm{j}}\frac{{2\pi }}{\lambda }{\bf{q}}_1^{\mathsf T}{{\bf{r}}_k}}},{{\rm{e}}^{{\rm{j}}\frac{{2\pi }}{\lambda }{\bf{q}}_2^{\mathsf T}{{\bf{r}}_k}}}, \cdots ,{{\rm{e}}^{{\rm{j}}\frac{{2\pi }}{\lambda }{\bf{q}}_N^{\mathsf T}{{\bf{r}}_k}}}} \right]^{\mathsf T}} \in {\mathbb{C}^{N \times 1}},
\end{equation}
where $\lambda$ is the wavelength and ${{\bf{r}}_k} = {\left[ {{u_k},{v_k}} \right]^{\mathsf T}}$ denotes the spatial AoA coordinate of the $k$-th target with ${u_k} = \cos {\phi _k}\sin {\theta _k} \in [-u_{\max},u_{\max}]$ and ${v_k} = \cos {\theta _k}\in [-v_{\max},v_{\max}]$ for $1 \le k \le K$. $\theta_k$ and $\phi _k$ are the physical elevation and azimuth AoAs of the LoS path from the $k$-th target to the receiver, respectively. For convenience, we denote the collection of $K$ targets' spatial AoA coordinates as $\tilde{{\bf r}} =\{ {{\bf{r}}_k},1\le k \le K\}$. Then, the received signal at the $t$-th snapshot for $1 \le t \le T$ is given by \cite{stoica1989music}
\begin{equation}\small
{\bf{y}}(t) = \mathop \sum \limits_{k = 1}^K {\bf{a}}\left( \tilde{{\bf q}},{{\bf{r}}_k} \right){s_k}(t) + {\bf{z}}(t) \in {\mathbb C^{N \times 1}},
\end{equation}
where ${s_k}(t)$ is the complex-valued signal incorporating path loss effect at the $t$-th snapshot of the $k$-th target. ${\bf{z}}(t) \sim \mathcal{CN}\left({\bf{0}}_{N \times 1},{\sigma^2}{\bf{I}}_N\right)$ represents the additive white Gaussian noise (AWGN) vector at the receiver, where $\sigma^2$ is the average noise power. To estimate the spatial AoAs of multiple targets, the received signals over $T$ snapshots can be given in the following matrix form as
\begin{equation}\label{signal}\small
{\bf{Y}} = {\bf{AS}} + {\bf{Z}} \in {\mathbb C^{N \times T}},
\end{equation}
with
\begin{equation}\small
{\bf{Y}} = [{\bf{y}}(1),{\bf{y}}(2), \cdots ,{\bf{y}}(T)]\in \mathbb C^{N \times T},
\end{equation}
\begin{equation}\small
{\bf{A}} = \left[ {\bf{a}}\left( {\tilde{{\bf q}}},{{\bf{r}}_1} \right),{\bf{a}}\left( {\tilde{{\bf q}}},{{\bf{r}}_2} \right), \cdots, {\bf{a}}\left( {\tilde{{\bf q}}},{{\bf{r}}_K} \right) \right] \in {\mathbb C^{N \times K}},
\end{equation}
\begin{equation}\small
{\bf{S}} = \left[ {{{\bf{s}}_1}}, {{{\bf{s}}_2}},\cdots, {{{\bf{s}}_K}} \right]^{\mathsf T} \in {{\mathbb C}^{K \times T}},
\end{equation}
\begin{equation}\small
{{\bf{s}}_k} = {[{s_k}(1),{s_k}(2), \cdots ,{s_k}(T)]^{\mathsf T}} \in {\mathbb C^{T \times 1}},
\end{equation}
and
\begin{equation}\small
{\bf{Z}} = \left[ {{{\bf{z}}(1)}}, {{{\bf{z}}(2)}},\cdots, {{{\bf{z}}(T)}} \right] \in {{\mathbb C}^{N \times T}}.
\end{equation}

\subsection{Multi-target AoA Estimation} \label{section:s22}
For any given antenna positions ${\tilde{{\bf q}}}$, we adopt the multiple signal classification (MUSIC) algorithm for estimating the spatial AoAs of multiple targets based on the received signals over $T$ snapshots \cite{schmidt1896music}. Specifically, the covariance matrix of the received signals in (\ref{signal}) is given by
\begin{equation}\small
{\bf{R}}_{\bf Y} = \mathbb E \{ {\bf Y}{\bf Y}^{\mathsf H} \} = {\bf{A}}{\bf{R}}_{\bf S} {\bf{A}}^{\mathsf H} + {\sigma^2}{\bf{I}}_N \in {\mathbb C}^{N \times N},
\end{equation}
where ${\bf{R}}_{\bf S} = {\bf{S}}{\bf{S}}^{\mathsf H} \in {\mathbb C}^{K \times K}$ denotes the covariance matrix of the source signals. Then, the singular value decomposition (SVD) of $\bf{R}_{\bf Y}$ can be obtained as
\begin{equation}\small
{\bf{R}}_{\bf Y} = \left[{\bf U}_{\rm s}, {\bf U}_{\rm z}\right]\left[\begin{array}{ll}
{\bf \Sigma}_{\rm s} & \\
& {\bf \Sigma}_{\rm z}
\end{array}\right]\left[{\bf U}_{\rm s}, {\bf U}_{\rm z}\right]^{\mathsf H},
\end{equation}
where ${\bf U}_{\rm s}\in \mathbb{C}^{N\times K}$ and ${\bf U}_{\rm z}\in \mathbb{C}^{N\times (N-K)}$ denote the singular vectors of the signal and noise subspaces, respectively. ${\bf \Sigma}_{\rm s} \in \mathbb{R}^{K\times K}$ and ${\bf \Sigma}_{\rm z} \in \mathbb{R}^{(N-K) \times (N-K)}$ are diagonal matrices with the diagonal elements denoting the singular values of the signal and noise subspaces, respectively. 
Then, the MUSIC algorithm based estimation of the spatial AoAs of multiple targets is to search for the maximum $K$ peaks of the following MUSIC spatial spectrum \cite{schmidt1896music}
\begin{equation}\small
P({\bar{\bf r}}_0) = \frac{1}{{\bf{a}}\left( \tilde{{\bf q}},{\bar{\bf r}}_0 \right)^{\mathsf H} {\bf U}_{\rm z} {\bf U}_{\rm z}^{\mathsf H}{\bf{a}}\left( \tilde{{\bf q}},{\bar{\bf r}}_0 \right)},
\end{equation}
where ${\bar{\bf r}}_0 \in[-u_{\max},u_{\max}] \times [-v_{\max},v_{\max}]$. The above estimate is asymptotically unbiased. For convenience, we assume that we have enough snapshots so that the above estimate is unbiased.

We denote the estimation of the spatial AoAs of the $k$-th target as ${\hat{\bf r}}_k$ for $1 \le k \le K$. Then, the MSE for multi-target AoA estimation can be given by
\begin{equation}\small
{\rm MSE} = \sum_{k=1}^K \mathbb E \{ \Vert{{\bf r}}_k - {\hat{\bf r}}_k \Vert_2^2 \},
\end{equation}
which is lower-bounded by the trace of the CRB matrix for multi-target AoA estimation, i.e.,
\begin{equation}\small
{\rm MSE} \ge {\rm tr} \left(\rm CRB(\tilde{{\bf q}}) \right).
\end{equation}

\subsection{CRB Characterization}
Let ${\bm{\omega}} = {[{u_1}, \cdots ,{u_K},{v_1}, \cdots ,{v_K}]^{\mathsf T}} \in {\mathbb R^{2K \times 1}}$ and ${\bm{\zeta }} = {\left[ {\Re {{\left\{ {{{\bf{s}}_1}} \right\}}^{\mathsf T}}, \cdots ,\Re {{\left\{ {{{\bf{s}}_K}} \right\}}^{\mathsf T}},\Im {{\left\{ {{{\bf{s}}_1}} \right\}}^{\mathsf T}}, \cdots ,\Im {{\left\{ {{{\bf{s}}_K}} \right\}}^{\mathsf T}}} \right]^{\mathsf T}} \in {\mathbb R^{2KT \times 1}}$ denote the unknown spatial AoAs and the real and imaginary parts of the complex signals to be estimated, respectively. Then, by vectorizing the received signals at the MA array in (\ref{signal}), we have
\begin{equation}\small
\begin{aligned}
{\rm{vec(}}{\bf{Y}})& = {\rm{vec(}}{\bf{AS}}) + {\rm{vec(}}{\bf{Z}})\\
&= ({{\bf{S}}^{\mathsf T}} \otimes {{\bf{I}}_N}){\rm{vec(}}{\bf{A}}) + {\rm{vec(}}{\bf{Z}}) \in {\mathbb C^{NT \times 1}},
\end{aligned}
\end{equation}
which follows complex Gaussian distribution with mean vector ${\bm \mu} = ({{\bf{S}}^{\mathsf T}} \otimes {{\bf{I}}_N}){\rm{vec(}}{\bf{A}})$ and covariance matrix ${\sigma^2}{\bf{I}}_{NT}$, i.e., ${\rm{vec(}}{\bf{Y}})  \sim \mathcal{CN}\left({\bm \mu},{\sigma^2}{\bf{I}}_{NT}\right)$. Then, the Fisher information matrix (FIM) ${\bf F} \in {\mathbb R^{(2K + 2KT) \times (2K + 2KT)}}$ for estimating $\bm \omega$ and $\bm \zeta$ can be expressed as
\begin{equation}\small
\begin{aligned}
{\bf{F}}& = \frac{2}{{{\sigma ^2}}}\left[ {\begin{array}{*{20}{c}}
	{\bf F}_{11}&{\bf F}_{12}\\
	{\bf F}_{21}&{\bf F}_{22}
	\end{array}} \right]\\
&\buildrel \Delta \over=\frac{2}{{{\sigma ^2}}}\left[ {\begin{array}{*{20}{c}}
	{\Re \left\{ {{\bf{D}}_{\bm{\omega }}^{\mathsf H}{{\bf{D}}_{\bm{\omega }}}} \right\}}&{\Re \left\{ {{\bf{D}}_{\bm{\omega }}^{\mathsf H}{{\bf{D}}_{\bm{\zeta }}}} \right\}}\\
	{\Re \left\{ {{\bf{D}}_{\bm{\zeta }}^{\mathsf H}{{\bf{D}}_{\bm{\omega }}}} \right\}}&{\Re \left\{ {{\bf{D}}_{\bm{\zeta }}^{\mathsf H}{{\bf{D}}_{\bm{\zeta }}}} \right\}}
	\end{array}} \right],
\end{aligned}
\end{equation}
where ${{\bf{D}}_{\bm{\omega }}} \in \mathbb C ^{NT \times 2K}$ is given by
\begin{equation}\small
{{\bf{D}}_{\bm{\omega }}} = \left[ {\frac{{\partial {\bm{\mu }}}}{{\partial {u_1}}}, \cdots ,\frac{{\partial {\bm{\mu }}}}{{\partial {u_K}}},\frac{{\partial {\bm{\mu }}}}{{\partial {v_1}}}, \cdots ,\frac{{\partial {\bm{\mu }}}}{{\partial {v_K}}}} \right],
\end{equation}
\begin{equation}\small
\begin{aligned}
\frac{{\partial {\bm{\mu }}}}{{\partial {u_k}}}& = ({{\bf{S}}^{\mathsf T}} \otimes {{\bf{I}}_N}){\rm{vec}}\left( {\frac{{\partial {\bf{A}}}}{{\partial {u_k}}}} \right)\\
& = ({{\bf{S}}^{\mathsf T}}{{\bf{e}}_k} \otimes {{\bf{I}}_N}) {\dot{\bf{a}}}_{\rm u}\left( {\tilde{{\bf q}},{{\bf{r}}_k}} \right), 1\le k \le K,
\end{aligned}
\end{equation}
\begin{equation}\small
\begin{aligned}
\frac{{\partial {\bm{\mu }}}}{{\partial {v_k}}}& = ({{\bf{S}}^{\mathsf T}} \otimes {{\bf{I}}_N}){\rm{vec}}\left( {\frac{{\partial {\bf{A}}}}{{\partial {v_k}}}} \right)\\
& = ({{\bf{S}}^{\mathsf T}}{{\bf{e}}_k} \otimes {{\bf{I}}_N}) {\dot{\bf{a}}}_{\rm v}\left( {\tilde{{\bf q}},{{\bf{r}}_k}} \right), 1\le k \le K,
\end{aligned}
\end{equation}
\begin{equation}\small
\begin{aligned}
{\dot{\bf{a}}}_{\rm u}\left( {\tilde{{\bf q}},{{\bf{r}}_k}} \right)&  \buildrel \Delta \over = \frac{{\partial {\bf{a}}\left( {\tilde{{\bf q}},{{\bf{r}}_k}} \right)}}{{\partial {u_k}}} \\
&= {\left[ {{\rm{j}}\frac{{2\pi }}{\lambda }{x_1}{{\rm{e}}^{{\rm{j}}\frac{{2\pi }}{\lambda }{\bf{q}}_1^{\mathsf T}{{\bf{r }}_k}}}, \cdots ,{\rm{j}}\frac{{2\pi }}{\lambda }{x_N}{{\rm{e}}^{{\rm{j}}\frac{{2\pi }}{\lambda }{\bf{q}}_N^{\mathsf T}{{\bf{r }}_k}}}} \right]^{\mathsf T}},
\end{aligned}
\end{equation}
\begin{equation}\small
\begin{aligned}
{\dot{\bf{a}}}_{\rm v}\left( {\tilde{{\bf q}},{{\bf{r}}_k}} \right)&  \buildrel \Delta \over = \frac{{\partial {\bf{a}}\left( {\tilde{{\bf q}},{{\bf{r}}_k}} \right)}}{{\partial {v_k}}} \\
&  = {\left[ {{\rm{j}}\frac{{2\pi }}{\lambda }{y_1}{{\rm{e}}^{{\rm{j}}\frac{{2\pi }}{\lambda }{\bf{q}}_1^{\mathsf T}{{\bf{r }}_k}}}, \cdots ,{\rm{j}}\frac{{2\pi }}{\lambda }{y_N}{{\rm{e}}^{{\rm{j}}\frac{{2\pi }}{\lambda }{\bf{q}}_N^{\mathsf T}{{\bf{r }}_k}}}} \right]^{\mathsf T}}.
\end{aligned}
\end{equation}
${{\bf{D}}_{\bm{\zeta }}} \in \mathbb C ^{NT \times 2KT}$ is given by
\begin{equation}\small
{{\bf{D}}_{\bm{\zeta }}} = \left[ {\frac{{\partial {\bm{\mu }}}}{{\partial \Re \{ {{\bf{s}}_1}\} }}, \cdots ,\frac{{\partial {\bm{\mu }}}}{{\partial \Re \{ {{\bf{s}}_K}\} }},\frac{{\partial {\bm{\mu }}}}{{\partial \Im \{ {{\bf{s}}_1}\} }}, \cdots ,\frac{{\partial {\bm{\mu }}}}{{\partial \Im \{ {{\bf{s}}_K}\} }}} \right],
\end{equation}
with 
\begin{equation}\small
\begin{aligned}
\frac{{\partial {\bm{\mu }}}}{{\partial \Re \{ {{\bf{s}}_k}\} }} = ({{\bf{I}}_T} \otimes {\bf{A}}){\rm{vec}}\left( {\frac{{\partial {\bf{S}}}}{{\partial \Re \{ {{\bf{s}}_k}\} }}} \right) = {{\bf{I}}_T} \otimes {\bf{a}}\left( {\tilde{{\bf q}},{{\bf{r}}_k}} \right),
\end{aligned}
\end{equation}
\begin{equation}\small
\begin{aligned}
\frac{{\partial {\bm{\mu }}}}{{\partial \Im \{ {{\bf{s}}_k}\} }} ={\rm j} ({{\bf{I}}_T} \otimes {\bf{A}}){\rm{vec}}\left( {\frac{{\partial {\bf{S}}}}{{\partial \Im \{ {{\bf{s}}_k}\} }}} \right) ={\rm j} {{\bf{I}}_T} \otimes {\bf{a}}\left( {\tilde{{\bf q}},{{\bf{r}}_k}} \right).
\end{aligned}
\end{equation}

Then, according to the derivation in Appendix \ref{app:A}, the CRB matrix for multi-target 2D spatial AoA estimation can be given by
\begin{equation}\label{crb}\small
\begin{aligned}
{\rm{CRB}}(\tilde{{\bf q}})=\frac{{{\sigma ^2}}}{2}{\left( {\Re \left\{ {({{\bf{1}}_{2}} \otimes {\bf{R}}_{\bf{S}}^{\mathsf T}) \odot {{\dot{\bf{ A}}}^{\mathsf H}}{\bf{\Pi }}_{\bf{A}}^ \bot \dot{\bf{ A}}} \right\}} \right)^{ - 1}},
\end{aligned}
\end{equation}
where ${\bf{\Pi }}_{\bf{A}}^ \bot = {{\bf{I}}_N} - {\bf{A}}{{({{\bf{A}}^{\mathsf H}}{\bf{A}})}^{ - 1}}{{\bf{A}}^{\mathsf H}} \in \mathbb C^{N\times N}$ is the orthogonal projection matrix of the column space of $\bf A$ and $\dot{\bf{ A}} = [{\dot{\bf{ A}}_{\rm{u}}},{\dot{\bf{ A}}_{\rm{v}}}]$ with
\begin{equation}\small
\begin{aligned}
{\dot{\bf{ A}}_{\rm{u}}} = \left[{\dot{\bf{a}}}_{\rm u}\left( {\tilde{{\bf q}},{{\bf{r}}_1}} \right), \cdots ,{\dot{\bf{a}}}_{\rm u}\left( {\tilde{{\bf q}},{{\bf{r}}_K}} \right)\right],
\end{aligned}
\end{equation}
\begin{equation}\small
\begin{aligned}
{\dot{\bf{ A}}_{\rm{v}}} = \left[{\dot{\bf{a}}}_{\rm v}\left( {\tilde{{\bf q}},{{\bf{r}}_1}} \right), \cdots ,{\dot{\bf{a}}}_{\rm v}\left( {\tilde{{\bf q}},{{\bf{r}}_K}} \right)\right].
\end{aligned}
\end{equation}

{\textbf{Remark}}: The CRB matrix for multi-target  one-dimensional (1D) spatial AoA estimation with 1D MA array is given by
\begin{equation}\small
\begin{aligned}
{{\rm{CRB}}}_{\rm u}(\tilde{{\bf q}}) =\frac{{{\sigma ^2}}}{2}{\left( {\Re \left\{ { {\bf{R}}_{\bf{S}}^{\mathsf T} \odot {{\dot{\bf{ A}}_{\rm u}}^{\mathsf H}}{\bf{\Pi }}_{\bf{A}_{\rm u}}^ \bot \dot{\bf{ A}}_{\rm u}} \right\}} \right)^{ - 1}},
\end{aligned}
\end{equation}
which can be regarded as a special case of the 2D CRB and its derivation is omitted due to the page limitation.

\subsection{Problem Formulation}
In practice, it is challenging to obtain the prior instantaneous AoA information of multiple targets for optimizing the MA array geometry. Nevertheless, the statistical distribution of the multi-target spatial AoAs can be exploited to improve sensing coverage performance. Therefore, to enhance the performance of multi-target spatial AoA estimation accuracy, we aim to minimize the expected sum of the CRBs for multi-target AoA estimation (i.e., the trace of the CRB matrix) by optimizing the positions of MAs. The associated optimization problem is formulated as
\begin{subequations}\label{opti}
\begin{align}\small
\mathop {{\rm{min}}}\limits_{ \tilde{{\bf q}} } \ & \mathbb E\left\{ {{\rm{tr}}({\rm{CRB}}(\tilde{{\bf q}}))} \right\}\label{opti:sub0}\\
{\rm{s}}.{\rm{t}}. \ & {\left\| {{{\bf{q}}_{n'}} - {{\bf{q}}_n}} \right\|_2} \ge {d_{\min }},1 \le n \ne n' \le N, \label{opti:sub1}\\
&{{\bf{q}}_n} \in {\cal C},1 \le n \le N,\label{opti:sub2}
\end{align}
\end{subequations}
where constraint (\ref{opti:sub1}) ensures the minimum distance ${d_{\min }}$ between any two antennas to avoid antenna coupling. Constraint (\ref{opti:sub2}) confines the antenna moving region. Since the expectation in the objective function is hard to be derived in a closed form and constraint (\ref{opti:sub1}) is non-convex, problem (\ref{opti}) is generally difficult to solve for the optimal solution. Therefore, we develop a swarm-based gradient descent algorithm to obtain a suboptimal solution in the next section.

\section{Proposed Solution} \label{section:s3}
In this section, we propose a swarm-based gradient descent algorithm to solve problem (\ref{opti}) suboptimally. To overcome the challenge that the objective function in (\ref{opti:sub0}) lacks a closed-form expression in general, we employ the Monte Carlo simulation method to approximate the objective function. In particular, given any statistical distribution of the targets, we randomly generate their angle samples and the corresponding transmitted signals over $M$ independent realizations. For a sufficient large $M$, the expectation of the trace of the CRB matrix for multi-target AoA estimation can be approximated by the sample average over all realizations, i.e.,
\begin{equation}\label{monte}\small
\begin{aligned}
{\mathbb E}\left\{ {{\rm{tr}}({\rm{CRB}}(\tilde{{\bf q}}))} \right\} &\approx \frac{1}{M}\sum\limits_{m = 1}^M {{\rm{tr}}({\rm{CR}}{{\rm{B}}_m(\tilde{{\bf q}})})}\\
& \buildrel \Delta \over = \frac{1}{M}\sum\limits_{m = 1}^M {{\psi _m}(\tilde{\bf{ q}})} \buildrel \Delta \over = \tilde \psi (\tilde{\bf{ q}}),
\end{aligned}
\end{equation}
where ${\rm{CR}}{{\rm{B}}_m(\tilde{{\bf q}})}$ denotes the CRB matrix in the $m$-th realization, $1 \le m \le M$. Then, the original problem (\ref{opti}) can be approximated as
\begin{subequations}\label{opti1}
	\begin{align}\small
	\mathop {{\min}}\limits_{ \tilde{{\bf q}} } \ & \tilde \psi (\tilde{\bf{ q}}) \label{opti1:object}\\
	{\rm{s}}.{\rm{t}}. \ & \text{\rm (\ref{opti:sub1}), (\ref{opti:sub2})}.
	\end{align}
\end{subequations}
To address the coupling of multiple antennas' positions in the objective function and constraint (\ref{opti:sub1}), we adopt the alternating optimization (AO) technique to iteratively optimize each antenna’s position with the others being fixed. We present the detailed algorithm for solving (\ref{opti1}) in the next subsection.

\subsection{CRB Optimization}
 For ease of optimizing the $n$-th antenna’s position, the objective function in (\ref{opti1:object}) is redefined as a function of ${\bf q}_n$, i.e., ${\tilde \psi _n}({{\bf{q}}_n})$. Then, the subproblem for optimizing ${\bf q}_n$ can be expressed as
 \begin{subequations}\label{opti2}
 	\begin{align}\small
 	\mathop {{\rm{min}}}\limits_{{{\bf{q}}_n}} \ & {{\tilde \psi }_n}({{\bf{q}}_n})\\
 	{\rm{s}}.{\rm{t}}. \ & {\left\| {{{\bf{q}}_n} - {{\bf{q}}_{n'}}} \right\|_2} \ge {d_{\min }},n' \ne n,1 \le n' \le N,\label{opti2:sub1}\\
 	&{{\bf{q}}_n} \in {\cal C},
 	\end{align}
 \end{subequations}
which is non-convex and can be efficiently solved by the swarm-based gradient descent algorithm \cite{swarm2024lu}, as detailed as follows.

In the swarm-based gradient descent algorithm, the swarm consists of $I$ agents, each of which is identified with an initial position, i.e., a candidate solution for problem (\ref{opti2}), ${{{\bf{q}}_n^{i,(0)}}}$, and an initial mass $g^{i,(0)} \in  (0,1]$ for $1 \le i \le I$. In the $\ell$-th iteration, the position of each agent is updated in the direction of the local gradient, which is given by
\begin{equation} \label{update}\small
{{{\bf{q}}_n^{i,(\ell+1)}}} = {\mathcal B} \left\{{{{\bf{q}}_n^{i,(\ell)}}} - \tau^{i,(\ell)} \nabla_{{{{\bf{q}}_n}}} {{\tilde \psi }_n}({\bf{q}}_n^{i,(\ell)})\right\},
\end{equation}
where ${\mathcal B}\left\{ {\bf{q}}_n  \right\}$ is a projection function that ensures the position of the $n$-th antenna to be in the confined moving region and is given by
\begin{equation}\small
[\mathcal{B}({\bf{q}}_n)]_j= \begin{cases}-\frac{A}{2}, & \text { if }[{\bf{q}}_n]_j<-\frac{A}{2}, \\ \frac{A}{2}, & \text { if }[{\bf{q}}_n]_j>\frac{A}{2}, \\ {[{\bf{q}}_n]_j,} & \text { otherwise}.\end{cases}
\end{equation}
$\tau^{i,(\ell)}$ is the step size depending on the relative mass of the $i$-th agent, which can be obtained via the backtracking line search \cite{swarm2024lu} and will be specified later. Specifically, an agent with a smaller objective value is of heavier mass, while an agent with a larger objective value is of lighter mass. Accordingly, heavier agents prioritize exploitation via smaller steps and local convergence; Lighter agents prioritize exploration via larger steps, expanding the search beyond local basins and strengthening global optimization. $\nabla_{{{{\bf{q}}_n}}} {{\tilde \psi }_n}({\bf{q}}_n^{i,(\ell)})$ is the gradient of the objective function at the local point ${\bf{q}}_n^{i,(\ell)}$, which can be numerically calculated according to the definition.

To obtain the step size in (\ref{update}), we first introduce the mass transition procedure of each agent in the iteration process. Let $\tilde \psi_{\max}^{(\ell)} = \max\limits_{i} {{\tilde \psi }_n}({\bf{q}}_n^{i,(\ell)})$ and $\tilde \psi_{\min}^{(\ell)} = \min\limits_{i} {{\tilde \psi }_n}({\bf{q}}_n^{i,(\ell)})$ denote the maximal and minimal objective values of the swarm in the $\ell$-th iteration, respectively. Accordingly, we denote the index of the agent that achieves the minimum objective value as $i_0 = \arg\min\limits_{i} {{\tilde \psi }_n}({\bf{q}}_n^{i,(\ell)})$. Then, the mass of the $i$-th agent is dynamically adjusted according to
 \begin{subequations}\label{mass}
	\begin{align}\small
	&g^{i,(\ell+1)} = g^{i,(\ell)} - \kappa^{i,(\ell)}g^{i,(\ell)}, i \neq i_0,\\
	&g^{i_0,(\ell+1)} = g^{i_0,(\ell)} + \sum_{i=1,i\neq i_0}^{I}\kappa^{i,(\ell)}g^{i,(\ell)},
	\end{align}
\end{subequations}
where $\kappa^{i,(\ell)} \in (0,1]$ is a parameter measuring the reduction of the mass of the $i$-th agent to the current global minimizer, i.e., the $i_0$-th agent, and can be obtained via
\begin{equation}\small
\kappa^{i,(\ell)} = \left(\frac{{{\tilde \psi }_n}({\bf{q}}_n^{i,(\ell)}) - \tilde \psi_{\min}^{(\ell)}}{\tilde \psi_{\max}^{(\ell)} - \tilde \psi_{\min}^{(\ell)}}\right)^p,
\end{equation}
where $p > 0$ is a fine-tuning parameter. As such, the total mass of the swarm gradually concentrates with the agents that are most likely to reach the global minimum of the solution space explored so far by the swarm. Then, the relative mass of the $i$-th agent is given by
\begin{equation}\small
{\tilde g}^{i,(\ell+1)} = \frac{g^{i,(\ell+1)}}{\max\limits_{i} g^{i,(\ell+1)}},1\le i\le I.
\end{equation}

Subsequently, the step size is initialized as a large positive value, $\tau^{i,(\ell)} = \tau_{\max}$. Then, we gradually shrink it by a factor $\varsigma \in (0,1)$, i.e., $\tau^{i,(\ell)} \leftarrow \varsigma\tau^{i,(\ell)}$, until the minimum inter-antenna distance constraint and the Armijo–Goldstein condition are both satisfied, i.e.,
\begin{equation} \label{ag}\small
 {{\tilde \psi }_n}({\bf{q}}_n^{i,(\ell+1)}) \le  {{\tilde \psi }_n}({\bf{q}}_n^{i,(\ell)}) - \xi \beta^{i,(\ell)}\tau^{i,(\ell)}{\left\|\nabla_{{{{\bf{q}}_n}}} {{\tilde \psi }_n}({\bf{q}}_n^{i,(\ell)}) \right\|_2^2},
\end{equation}
where $\xi \in (0,1)$ is a predefined parameter to control the decreasing speed of the objective function. $\beta^{i,(\ell)} $ is a parameter related to the relative mass of the $i$-th agent, which is given by
\begin{equation}\small
\beta^{i,(\ell)} = ({\tilde g}^{i,(\ell+1)})^q,
\end{equation}
with $q > 0$ allowing fine-tuning of the dependence on the relative mass.

We summarize the overall algorithm for solving problem (\ref{opti1}) in Algorithm \ref{alg1}. Specifically, in line 1, we initialize the antenna positions $\tilde{{\bf q}}$ to form a uniform planar array (UPA) with full aperture. Next, we generate $M$ independent target angle samples and the corresponding transmitted signals, which are then used to calculate the approximate objective function in (\ref{monte}). Then, in lines 4-24, we iteratively optimize the position of each antenna using the swarm-based gradient algorithm, where $J$ represents the maximum iteration number for AO and $L$ denotes the maximum iteration number for the swarm-based gradient descent algorithm. It is noted that the position of the agent with the minimum objective value among all the agents in the swarm will be output as the obtained suboptimal solution at the end of the inner iteration. The outer iteration for AO and the inner iteration for swarm-based gradient descent optimization terminate if the relative decrease of the objective function between two consecutive iterations is no larger than a predefined threshold $\varepsilon$ or the maximum iteration number is attained.

The computational complexity of Algorithm \ref{alg1} is analyzed as follows. The computation of the expectation of the trace of the CRB matrix in (\ref{monte}) entails a complexity of $\mathcal{O}\left(MK^2(NT+K) \right)$. The local gradient calculation in (\ref{update}) involves $2N$ times of computing the expectation of the trace of the CRB matrix in (\ref{monte}). Thus, the corresponding computational complexity is given by $\mathcal{O}\left(MNK^2(NT+K) \right)$. By denoting the maximum number of backtracking line search in lines 14-17 as $B$, the corresponding computational complexity is $\mathcal{O}\left(BMNK^2(NT+K) \right)$. Given the maximum iteration number for AO, $J$, and the maximum iteration number for the swarm-based gradient descent algorithm, $L$, the worst-case computational complexity of Algorithm \ref{alg1} for solving (\ref{opti1}) is thus given by $\mathcal{O}\left(JLBMNK^2(NT+K) \right)$. It is worth noting that the considered sensing coverage problem is optimized based on statistical distribution of targets, which does not require frequent movement of antennas. Once the antennas have been moved to the optimized positions, the array geometry remains unchanged for a long period unless the coverage requirement changes. Therefore, the computational complexity of the proposed algorithm is acceptable.
\begin{algorithm}[!tb]\small
	\caption{Swarm-based gradient descent algorithm for problem (\ref{opti1})}
	\label{alg1}
	\begin{algorithmic}[1]
		\REQUIRE
		$K,N,T,I,A,\sigma^2,\lambda,d_{\min},p,q,\tau_{\max},\varsigma,\xi,J,L,\varepsilon$.
		\ENSURE
		$\tilde{{\bf q}}$.
		\STATE
		Initialize the antenna positions $\tilde{{\bf q}}$.
		\STATE
		Generate $M$ independent target angle samples and corresponding transmitted signals.
		\STATE
		Obtain $\tilde \psi (\tilde{\bf{ q}})$ according to (\ref{monte}).
		\STATE Set $j = 1$.
		\REPEAT
		\FOR{$n = 1$ to $N$}
		\REPEAT
		\STATE Set $\ell = 0$.
		\STATE Initialize ${\bf{ q}}_n^{1,(0)} = {\bf{q}}_n$ and initialize ${\bf{ q}}_n^{i,(0)}$ randomly while satisfying constraint (\ref{opti2:sub1}) for $2 \le i \le I$.
		\STATE Initialize $g^{i,(0)} = \frac{1}{I}$ for $1 \le i \le I$.
		\STATE Update the masses of all the agents via (\ref{mass}).
		\FOR{$i=1$ to $I$}
		\STATE Initialize $\tau^{i,(\ell)} \leftarrow \tau_{\max}$.
		\REPEAT
		\STATE Update $\tau^{i,(\ell)} \leftarrow \varsigma\tau^{i,(\ell)}$.
		\STATE Update the $i$-th agent's position via (\ref{update}).
		\UNTIL constraint (\ref{opti2:sub1}) and (\ref{ag}) are both satisfied.
		\ENDFOR
		\STATE Update $\ell \leftarrow \ell+1$.
		\STATE Update ${\bf{q}}_n \leftarrow {\bf{ q}}_n^{i_0,(\ell)}$, $\tilde \psi (\tilde{\bf{ q}}) \leftarrow \tilde \psi_{\min}^{(\ell)}$.
		\UNTIL The relative decrease of the objective function is no larger than $\varepsilon$ or the maximum iteration number $L$ is attained.
		\ENDFOR
		\STATE Update $j \leftarrow j+1$.
		\UNTIL The relative decrease of the objective function is no larger than $\varepsilon$ or the maximum iteration number $J$ is attained.
		\RETURN $\tilde{{\bf q}}$.
	\end{algorithmic}
\end{algorithm}
\vspace{-1em}
\subsection{Performance Lower-bound Analysis}
To demonstrate the superiority of our proposed design, we derive the performance lower-bound on the objective function $\tilde \psi (\tilde{\bf{ q}})$ in problem (\ref{opti1}). First, let $\tilde{\bf{x}} = [x_1,x_2,\cdots,x_N]^{\mathsf T}$ and $\tilde{\bf{y}} = [y_1,y_2,\cdots,y_N]^{\mathsf T}$, corresponding to the antenna location parameters $\tilde{\bf{ q}}$. The variance function is then defined as ${\rm{var}}(\tilde{\bf{x}}) = \frac{1}{N}\sum\nolimits_{n=1}^{N}(x_n - \mu(\tilde{\bf{x}}))^2$ with $\mu(\tilde{\bf{x}}) = \frac{1}{N}\sum \nolimits_{n=1}^{N} x_n$, and ${\rm{var}}(\tilde{\bf{y}})$ and $\mu(\tilde{\bf y})$ are similarly defined. The covariance function is defined as ${{\rm cov}} {{(\tilde{ \bf{x}},\tilde{\bf{y}})}} = \frac{1}{N}\sum \nolimits_{n=1}^{N} (x_n - \mu(\tilde{\bf{x}}))(y_n - \mu(\tilde{\bf{y}}))$. Additionally, we denote the orthogonal projection matrix of the space spanned by ${\bf{a}}\left( {\tilde{\bf{q}},{{\bf{r}}_k}}\right)$ as
\begin{equation}\small
{\bf{\Pi }}_k^ \bot = {\bf I}_N- { \frac{{{{\bf{a}}\left( {\tilde{\bf{q}},{{\bf{r}}_k}} \right)}{\bf{a}}\left( {\tilde{\bf{q}},{{\bf{r}}_k}} \right)^{\mathsf H}}}{{{{\left\| {{{\bf{a}}\left( {\tilde{\bf{q}},{{\bf{r}}_k}} \right)}} \right\|}^2_2}}}},1 \le k \le K.
\end{equation}
Furthermore, we define
\begin{equation}\small
\zeta ^*_{\iota,k} = \frac{{\Re \left\{ {\dot{\bf{ a}}_{\rm{u}}^{\mathsf H}\left( {\tilde{\bf{q}},{{\bf{r}}_k}} \right){\bf{\Pi }}_k^ \bot {{\dot{\bf{ a}}}_{\rm{v}}}\left( {\tilde{\bf{q}},{{\bf{r}}_k}} \right)} \right\}}}{{\left\| {{\bf{\Pi }}_k^ \bot {{\dot{\bf{ a}}}_{\iota}}\left( {\tilde{\bf{q}},{{\bf{r}}_k}} \right)} \right\|_2^2}}, \iota \in \left\{ \rm u,v \right\},1 \le k \le K.
\end{equation}
Then, the lower-bound on the objective function $\tilde \psi (\tilde{\bf{ q}})$ in problem (\ref{opti1}) can be obtained via the following theorem.
\begin{theorem} \label{theorem}
The objective function $\tilde \psi (\tilde{\bf{ q}})$ in (\ref{opti1}) is lower-bounded by
\begin{equation} \label{lower} \small
\begin{aligned}
\tilde \psi (\tilde{\bf{ q}}) &\buildrel {\text{(\rm a)}} \over \ge \frac{K{{\sigma ^2}{\lambda ^2}}}{{8NTP_{\rm s}{\pi ^2}}} {\left( {\frac{1}{{{\rm{var}}(\tilde{\bf{x}}) - \frac{{{{\rm cov}} {{(\tilde{ \bf{x}},\tilde{\bf{y}})}^2}}}{{{\rm{var}}(\tilde{\bf{y}})}}}} + \frac{1}{{{\rm{var}}(\tilde{\bf{y}}) - \frac{{{{\rm cov}} {{(\tilde{\bf{x}},\tilde{\bf{y}})}^2}}}{{{\rm{var}}(\tilde{\bf{x}})}}}}} \right)}\\
&\buildrel {\text{(\rm b)}} \over \ge \frac{K{{\sigma ^2}{\lambda ^2}}}{{NTP_{\rm s}A^2}{\pi ^2}},
\end{aligned}
\end{equation}
where we assume that the signal energies from targets are the same to facilitate the theoretical analysis, i.e., $P_{\rm s} = {{\left\| {{{\bf{s}}_k}} \right\|_2^2}}/{T}$, $1 \le k \le K$. It is noted that the equality at (a) in (\ref{lower}) holds if and only if, for any $\iota, \iota' \in \left\{ \rm u,v \right\}$,
\begin{equation}\label{condition1}\small
\Re \left\{ {{{\left[ {{\bf{R}}_{\rm{s}}^{\mathsf T}} \right]}_{k,k'}}{{\dot{\bf{ a}}}_{\iota}}{{\left( {\tilde{\bf{q}},{{\bf{r}}_k}} \right)}^{\mathsf H}}{\bf{\Pi }}_{\bf{A}}^ \bot{{\dot{\bf{ a}}}_{\iota'}}\left( {\tilde{\bf{q}},{{\bf{r}}_{k'}}} \right)} \right\} = 0,k \ne k',
\end{equation}
and, for any $\iota \neq \iota' \in \left\{ \rm u,v \right\}$,
\begin{equation}\label{condition2}\small
	{\bf A}^{\mathsf H}{\bf{\Pi }}_k^ \bot \left( {{{\dot{\bf{ a}}}_{\iota}}\left( {{\bf{q}},{{\bf{r}}_k}} \right) - \zeta ^*_{\iota,k}{{\dot{\bf{ a}}}_{\iota'}}\left( {{\bf{q}},{{\bf{r}}_k}} \right)} \right) = {\bf{0}}_{K \times 1},
\end{equation}
for $1 \le k \neq k' \le K$. Furthermore, the equality at (b) holds if and only if
\begin{subequations}\label{condition3}
	\begin{align}\small
	{{{\rm cov}} {{(\tilde{ \bf{x}},\tilde{\bf{y}})}}} &= 0,\\
	{{\rm{var}}(\tilde{\bf{x}})} &= {{\rm{var}}(\tilde{\bf{y}})},\\
	\mu(\tilde{\bf{x}}) &= \mu(\tilde{\bf{y}}) = 0,\\
	x_n^2 + y_n^2 &= \frac{A^2}{2},1\le n \le N.
	\end{align}
\end{subequations}
\end{theorem}
\begin{IEEEproof}[Proof]
	Please refer to Appendix \ref{app:B}.
\end{IEEEproof}

Theorem \ref{theorem} indicates that the equality at (a) always holds for the single-target case, which is consistent with the result in \cite{ma2024sensing}. For the case of multiple targets, Theorem \ref{theorem} shows that if (43) and (44) are both satisfied, the trace of the CRB matrix for joint multi-target AoA estimation is equivalent to the sum of traces of the CRB matrices for estimating each target's AoAs separately as derived in \cite{ma2024sensing}, where the AoA estimation of one target is critically independent of the AoAs of all other targets. Specifically, condition (\ref{condition1}) requires the interference between estimating AoAs of different targets to be zero and condition (\ref{condition2}) guarantees the maximum sensing sensitivity for estimating the 2D AoAs of each target. In light of this, to approach the lower-bound on the sum of the CRBs for multi-target AoA estimation, on one hand, it is desirable to reduce the correlation between the waveforms in the time domain for different targets. As such, condition (\ref{condition1}) is satisfied if the time-domain waveforms for different targets are orthogonal. On the other hand, we should optimize the antenna array geometry in the spatial domain to decrease the normalized correlation of sensitivity vectors for estimating the 2D AoAs of different
targets. Specifically, the sensitivity vector is defined as ${\bf{\Pi }}_{\bf{A}}^ \bot{{\dot{\bf{ a}}}_{\iota }}{\left( {\tilde{\bf{q}},{{\bf{r}}_{k}}} \right)}$ for $ 1\le k\le K,\iota\in \left\{ \rm u,v \right\}$, representing the projection of the derivative of the steering vector of each target w.r.t. its angular position into the noise subspace. Then, the normalized correlation of sensitivity vectors is given by
\begin{equation}\small\label{corr}
\begin{aligned}
\rho_{k,k'}(\iota,\iota') = {\frac{{\left| {{{\dot{\bf{ a}}}_{\iota }}{{\left( {\tilde{\bf{q}},{{\bf{r}}_{k}}} \right)}^{\mathsf H}}{\bf{\Pi }}_{\bf{A}}^ \bot {{\dot{\bf{ a}}}_{\iota'}}\left( {\tilde{\bf{q}},{{\bf{r}}_{k'}}} \right)} \right|}^2}{{{{\left\| {{\bf{\Pi }}_{\bf{A}}^ \bot {{\dot{\bf{ a}}}_{\iota}}\left( {\tilde{\bf{q}},{{\bf{r}}_{k}}} \right)} \right\|}_2^2}{{\left\| {{\bf{\Pi }}_{\bf{A}}^ \bot {{\dot{\bf{ a}}}_{\iota'}}\left( {\tilde{\bf{q}},{{\bf{r}}_{k'}}} \right)} \right\|}_2^2}}}},
\end{aligned}
\end{equation}
for $ 1\le k \neq k' \le K,\iota,\iota'\in \left\{ \rm u,v \right\}$, which is an important parameter measuring the interference between multi-target AoA estimation. In other words, a smaller value of $\rho_{k,k'}(\iota,\iota')$ means that the estimation of one target's AoA is less impacted by that of another one. It is noted that condition (\ref{condition1}) is satisfied if (\ref{corr}) is reduced to zero. Moreover, to decrease the sum of the CRBs for multi-target sensing, we should increase the effective power of sensitivity vectors incorporating the interference between the estimation of the each target's 2D AoAs, which are defined as
\begin{equation}\small\label{sensi1}
\begin{aligned}
\omega_{{\rm u},k} &=\mathop{\min }\limits_{\zeta_{{\rm u},k} \in \mathbb R} {\left\| {{\bf{\Pi }}_{\bf{A}}^ \bot \left( {{{\dot{\bf{ a}}}_{\rm{u}}}\left( {\tilde{\bf{q}},{{\bf{r}}_k}} \right) - \zeta_{{\rm u},k} {{\dot{\bf{ a}}}_{\rm{v}}}\left( {\tilde{\bf{q}},{{\bf{r}}_k}} \right)} \right)} \right\|}_2^2\\
&= {\left\| {{\bf{\Pi }}_{\bf{A}}^ \bot {{\dot{\bf{ a}}}_{\rm{u}}}\left( {\tilde{\bf{q}},{{\bf{r}}_k}} \right)} \right\|_2^2 - \frac{{\Re {{\left\{ {\dot{\bf{ a}}_{\rm{u}}^{\mathsf H}\left( {\tilde{\bf{q}},{{\bf{r}}_k}} \right){\bf{\Pi }}_{\bf{A}}^ \bot {{\dot{\bf{ a}}}_{\rm{v}}}\left( {\tilde{\bf{q}},{{\bf{r}}_k}} \right)} \right\}}^2}}}{{{{\left\| {{\bf{\Pi }}_{\bf{A}}^ \bot {{\dot{\bf{ a}}}_{\rm{v}}}\left( {\tilde{\bf{q}},{{\bf{r}}_k}} \right)} \right\|}_2^2}}}},
\end{aligned}
\end{equation}
and
\begin{equation}\small\label{sensi2}
\begin{aligned}
\omega_{{\rm v},k} &=\mathop{\min }\limits_{\zeta_{{\rm v},k} \in \mathbb R} {\left\| {{\bf{\Pi }}_{\bf{A}}^ \bot \left( {{{\dot{\bf{ a}}}_{\rm{v}}}\left( {\tilde{\bf{q}},{{\bf{r}}_k}} \right) - \zeta_{{\rm v},k} {{\dot{\bf{ a}}}_{\rm{u}}}\left( {\tilde{\bf{q}},{{\bf{r}}_k}} \right)} \right)} \right\|}_2^2\\
&= {\left\| {{\bf{\Pi }}_{\bf{A}}^ \bot {{\dot{\bf{ a}}}_{\rm{v}}}\left( {\tilde{\bf{q}},{{\bf{r}}_k}} \right)} \right\|_2^2 - \frac{{\Re {{\left\{ {\dot{\bf{ a}}_{\rm{u}}^{\mathsf H}\left( {\tilde{\bf{q}},{{\bf{r}}_k}} \right){\bf{\Pi }}_{\bf{A}}^ \bot {{\dot{\bf{ a}}}_{\rm{v}}}\left( {\tilde{\bf{q}},{{\bf{r}}_k}} \right)} \right\}}^2}}}{{{{\left\| {{\bf{\Pi }}_{\bf{A}}^ \bot {{\dot{\bf{ a}}}_{\rm{u}}}\left( {\tilde{\bf{q}},{{\bf{r}}_k}} \right)} \right\|}_2^2}}}},
\end{aligned}
\end{equation}
for $1\le k \le K$. The effective power of sensitivity vector measures the sensibility of the estimator to small changes of AoAs, the increase of which helps improve the sensing accuracy for each individual target. Specifically, as proved in Appendix \ref{app:B}, (\ref{condition2}) is the sufficient and necessary condition for maximization of (\ref{sensi1}) and (\ref{sensi2}). Therefore, condition (\ref{condition2}) is satisfied if both (\ref{sensi1}) and (\ref{sensi2}) achieve their maximum. While it is generally challenging to satisfy conditions (\ref{condition1}) and (\ref{condition2}) simultaneously, our proposed MA-based design is capable of striking a fine trade-off between reducing the normalized sensitivity vector correlation and increasing the effective power of sensitivity vectors in the pursuit of the minimum CRB for multi-target AoA estimation via antenna position optimization, compared to conventional FPA arrays and the single-target-oriented MA arrays in \cite{ma2024sensing}. This will be evaluated in the simulation section to demonstrate the superiority of our proposed design. Furthermore, condition (\ref{condition3}) suggests that it is desired to deploy MAs in a centrally symmetrical manner to further approach the lower-bound on the sum of the CRBs. In addition, the lower-bound on the sum of the CRBs is inversely proportional to the region size, i.e., $A^2$. Therefore, the expected sum of the CRBs can be efficiently reduced by increasing the size of the antenna movement region and carefully designing the array geometry.
\section{Numerical Results} \label{section:s4}

\begin{figure}[t]
	\centering
	\includegraphics[width= 7 cm]{{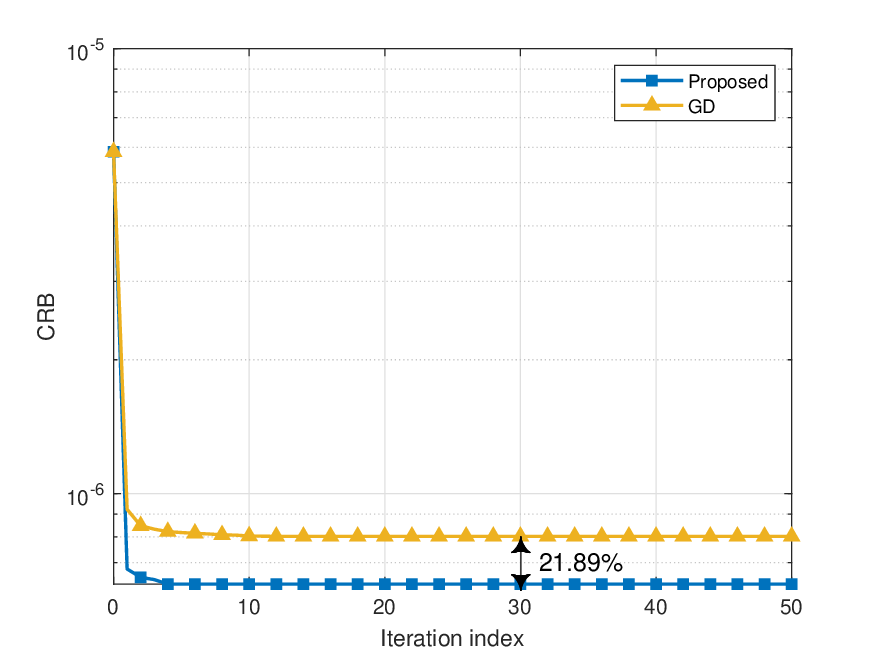}}
	\caption{Convergence performance of the proposed algorithm and the GD scheme.}
	\label{fig:Convergence}
	\vspace{-2em}
\end{figure}

\begin{figure}[t]
	\begin{center}
		\includegraphics[width= 7 cm]{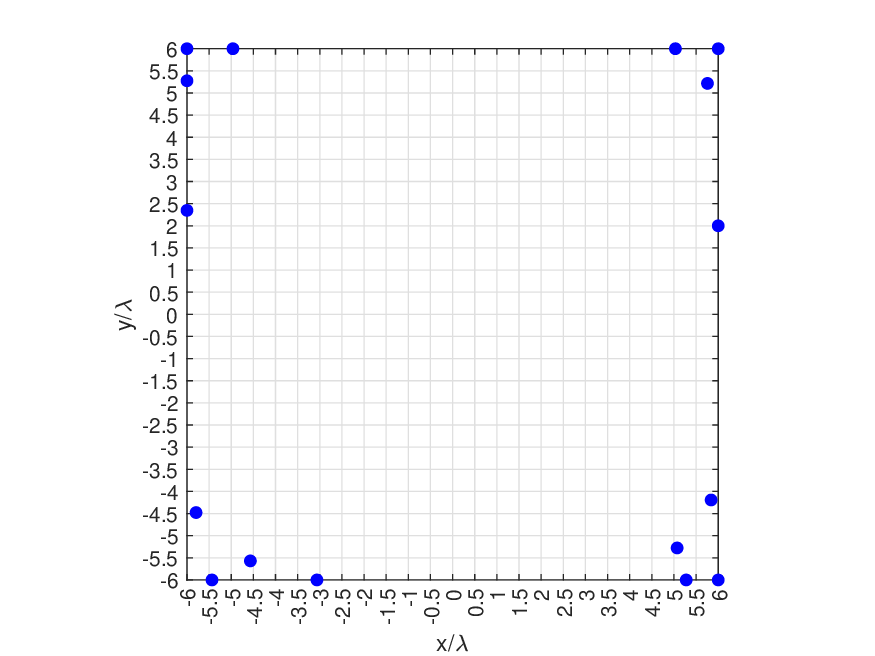}
		\caption{Illustration of the optimized MA array.}
		\label{fig:Geometry}
	\end{center}
	\vspace{-2em}
\end{figure}
\subsection{Simulation Setup and Benchmark Schemes}
In the simulations, we consider $K = 5$ targets that are randomly distributed within the spatial region $[-u_{\max},u_{\max}] \times [-v_{\max},v_{\max}]$ with $u_{\max} = v_{\max} = 0.6$. The number of MAs is set to $N = 16$, whose positions are confined in a square moving region with the size of $12\lambda \times 12\lambda$, i.e., $A = 12\lambda$, where the wavelength is set to $\lambda = 0.05$ m. The number of snapshots is set to $T = 64$. The average received SNR of each target's signals, defined as $P_{\rm s}/{\sigma^2}$, is 10 dB. The maximum iteration numbers for both the AO and the swarm-based gradient descent algorithm are set to $J=L=50$. The termination threshold in Algorithm \ref{alg1} is set to $\varepsilon = 10^{-3}$. The swarm-based gradient descent algorithm related parameters are set to $I = 25$, $p = 2$, $q = 0.5$, $\tau_{\max} = 0.25\lambda$, and $\xi = 0.6$ \cite{swarm2024lu}. The curves in the figures are the averaged results over 200 independent random target distributions.

To demonstrate the superiority of the proposed MA-based design, we compare its CRB performance, measuring the theoretical sensing performance limit, and MSE performance, evaluating the practical estimation accuracy, against the following benchmark schemes:
\begin{itemize}
	\item {\textit{\textbf{Lower-bound}}}: The lower-bound of the expected sum of the CRBs for multi-target AoA estimation is calculated via (\ref{lower}).
	
	\item {\textit{\textbf{MA array in \cite{ma2024sensing}}}}: The MA array geometry designed for single target sensing in \cite{ma2024sensing} is adopted, while the MUSIC algorithm is employed for multi-target sensing.
	
	\item {\textit{\textbf{Dense UPA}}}: The antenna elements are configured as an UPA with half-wavelength inter-antenna spacing.
	
	\item {\textit{\textbf{Sparse UPA}}}: The antenna elements are configured to form an UPA to achieve the largest aperture with the inter-antenna spacing of $A/(\lceil {\sqrt{N}} \rceil - 1)$ both horizontally and vertically. 
\end{itemize}
\vspace{-1.2em}
\subsection{Simulation Results}
We first evaluate the convergence performance of Algorithm \ref{alg1} in Fig. \ref{fig:Convergence}. It is shown that our proposed algorithm exhibits fast convergence, and the objective value remains nearly unchanged within 30 iterations, which verifies the effectiveness of the proposed antenna position optimization method. Moreover, it is also observed that the swarm-based gradient algorithm outperforms the conventional gradient ascent (GD) algorithm by 21.89\% thanks to the interplay between multiple agents. Furthermore, we illustrate the optimized MA array geometry in Fig. \ref{fig:Geometry}, which shows that the antenna elements tend to be deployed as far from the center of the square region as possible to enlarge the aperture of the MA array, thereby increasing the angular resolution for enhancing the multi-target AoA estimation performance. In addition, the antenna elements are non-uniformly configured within the movable region to help reduce the interference between the spatial AoA estimation of different targets.

%
%

\begin{figure*}[t]
	\centering
	\begin{minipage}[t]{0.3\textwidth}
		\centering
		{\includegraphics[width=1.05 \linewidth]{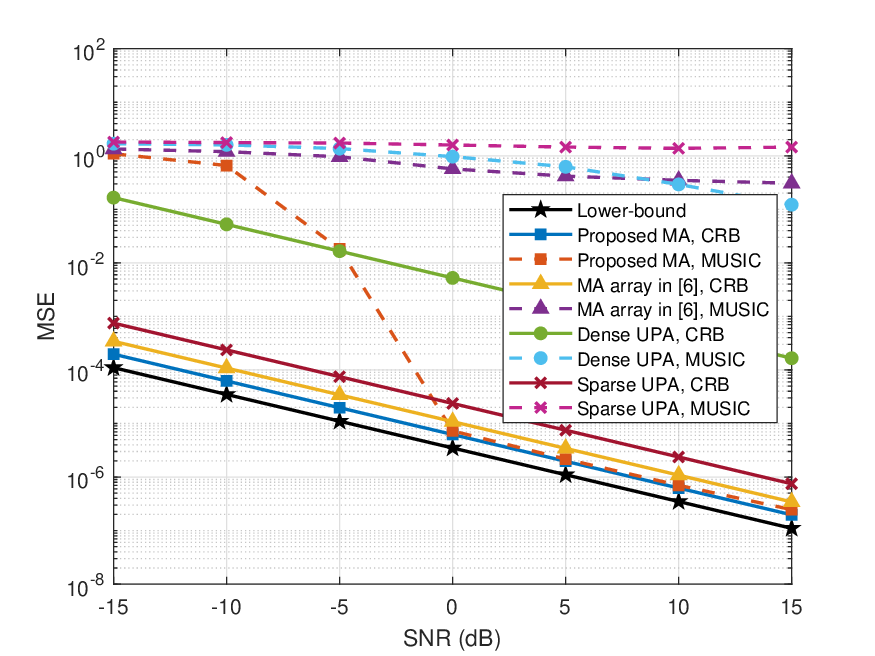}} 
		\caption{Performance comparison of different schemes versus received SNR.}
		\label{fig:SNR}
	\end{minipage}
\quad
	\begin{minipage}[t]{0.3\textwidth}
		\centering
		{\includegraphics[width=1.05 \linewidth]{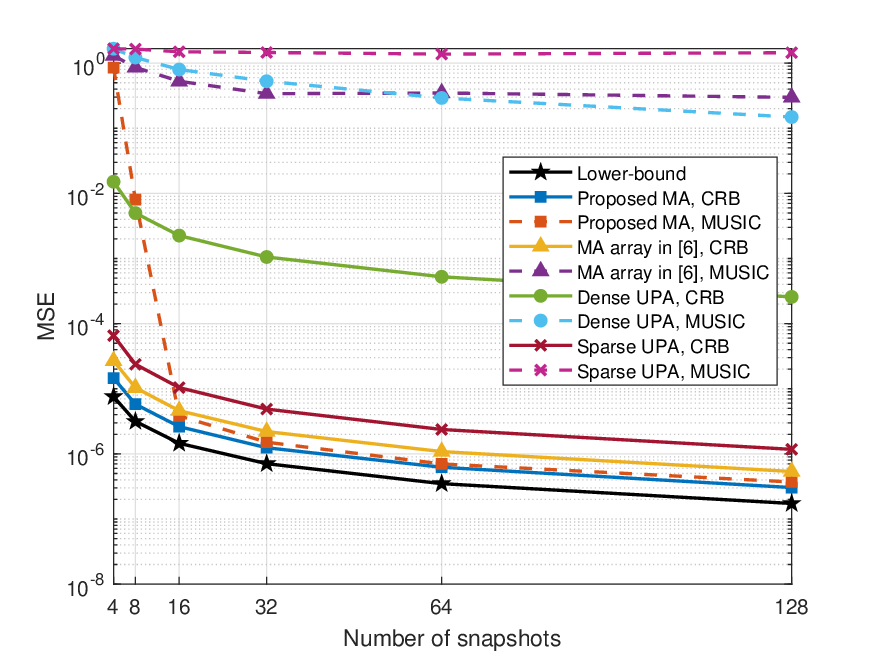}}  
		\caption{Performance comparison of different schemes versus number of snapshots.}
		\label{fig:SnapshotNumber}
	\end{minipage}
\quad
	\begin{minipage}[t]{0.3\textwidth}
	\centering
	\includegraphics[width=1.05 \linewidth]{{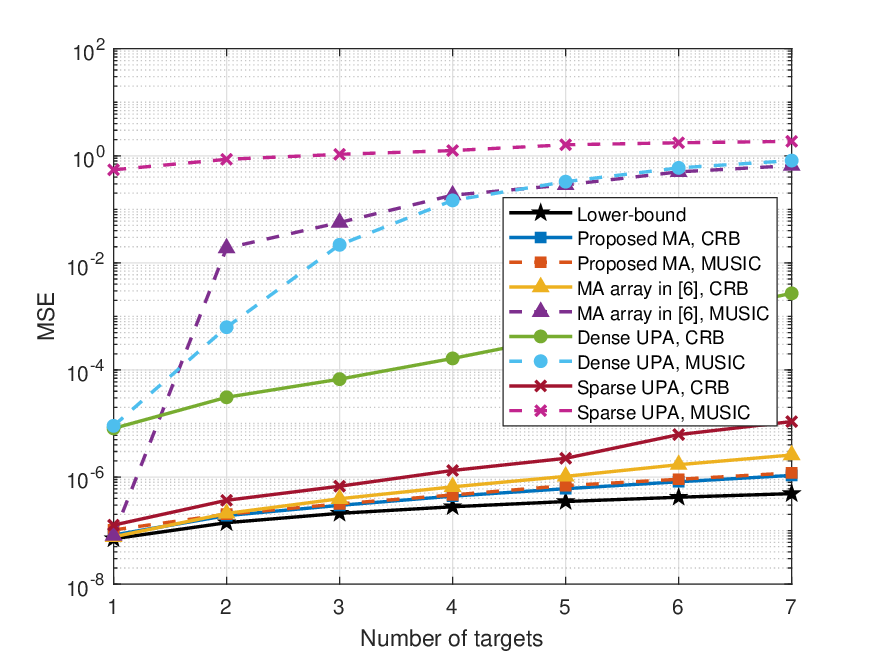}}
	\caption{Performance comparison of different schemes versus number of targets.}
	\label{fig:TargetNumber}
\end{minipage}
\vspace{-1em}
\end{figure*}

\begin{figure*}[t]
	\centering
	\begin{minipage}[t]{0.3\textwidth}
		\centering
		{\includegraphics[width=1.05 \linewidth]{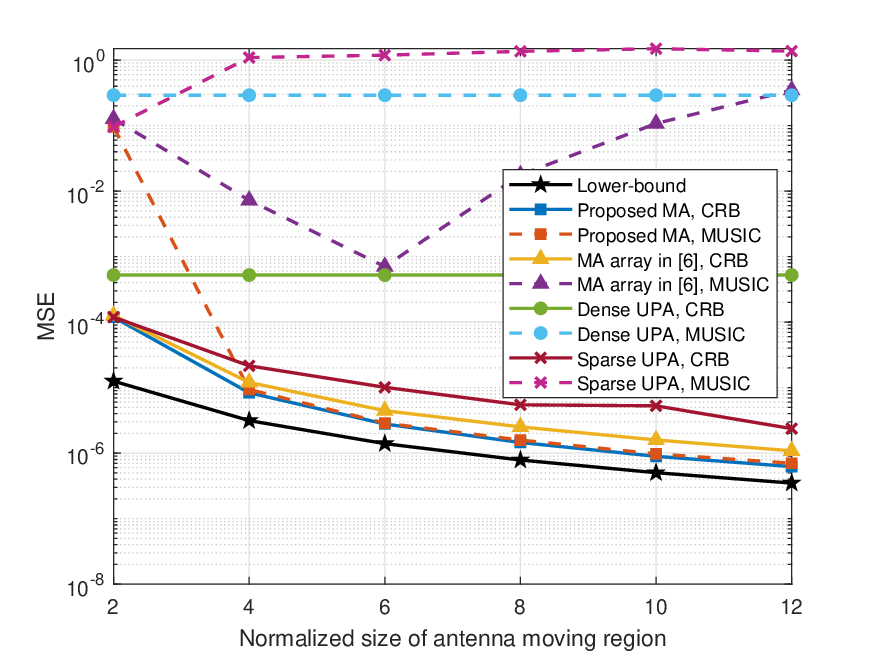}} 
		\caption{Performance comparison of different schemes versus normalized size of antenna moving region.}
		\label{fig:RegionSize}
	\end{minipage}
	\quad
	\begin{minipage}[t]{0.3\textwidth}
		\centering
		{\includegraphics[width=1.05 \linewidth]{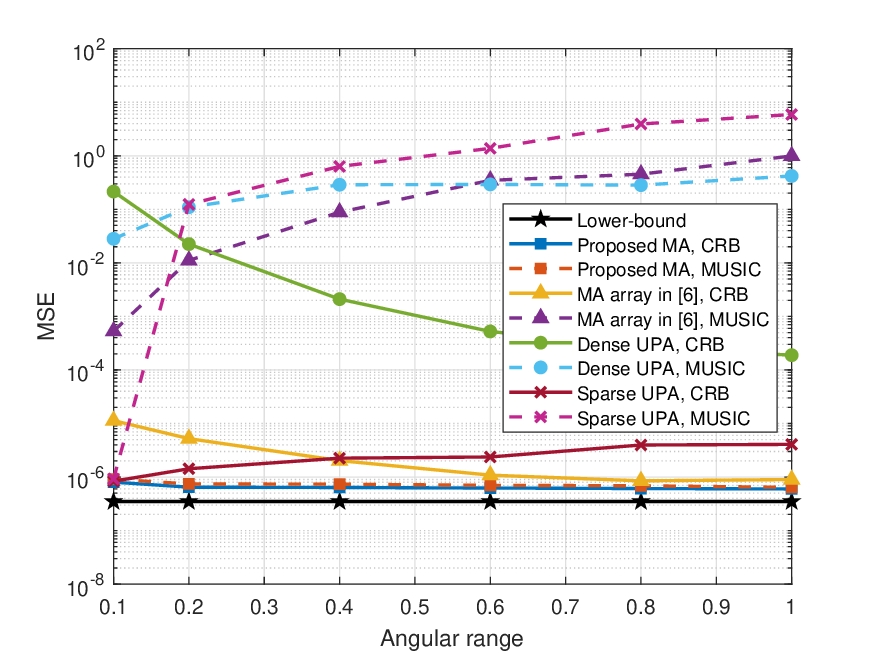}}  
		\caption{Performance comparison of different schemes versus angular range of target distribution.}
		\label{fig:AngleRange}
	\end{minipage}
	\quad
	\begin{minipage}[t]{0.3\textwidth}
		\centering
		{\includegraphics[width=1.05 \linewidth]{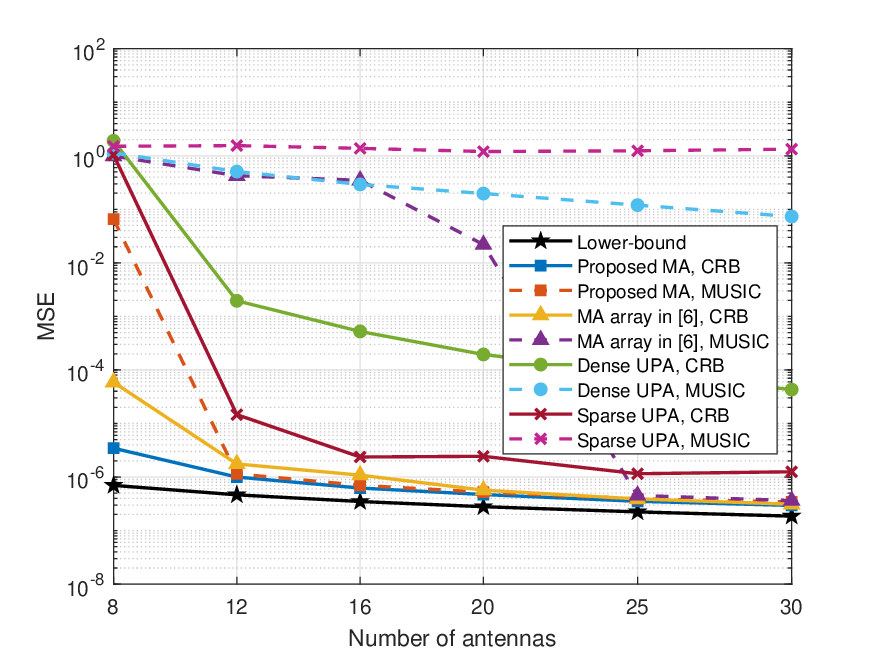}}  
		\caption{Performance comparison of different schemes versus number of antennas.}
		\label{fig:AntennaNumber}
	\end{minipage}
	\vspace{-1.5em}
\end{figure*}

In Fig. \ref{fig:SNR}, we compare the performance of the proposed and benchmark schemes versus the received SNR. It shows that the proposed scheme achieves superior CRB performance compared to benchmark schemes and is close to the performance lower-bound. It is also observed that the curve depicting AoA estimation MSE with the MUSIC algorithm of our proposed scheme approaches the CRB in the high-SNR regime. Additionally, our proposed MA-based design always achieves the lowest MSE performance compared to the benchmark schemes, which demonstrates the practicality of the proposed MA array for improving AoA estimation accuracy.

Similar results can also be observed from Fig. \ref{fig:SnapshotNumber}, where the proposed MA scheme achieves much lower CRBs and MSEs compared to the MA array in \cite{ma2024sensing} for single target sensing and conventional FPA-based systems. As the number of snapshots increases, the sensing performance experiences a certain degree of performance improvement. This is because more snapshots help accumulate higher signal power to mitigate the impact of environmental noise on the estimation performance.

In Fig. \ref{fig:TargetNumber}, we compare the performance of the proposed and benchmark schemes versus the number of targets. It is depicted that the CRB and MSE of all schemes increase with the number of targets. We can also observe that the performance gap between the proposed scheme and the performance lower-bound and that between the proposed scheme and the single-target-oriented MA array in \cite{ma2024sensing} become larger as the number of targets increases. The reason is that a larger number of targets entails greater interference between the estimation of different targets. Nevertheless, our proposed design achieves considerable performance gain by benefiting from flexible antenna position configuration to overcome this challenge. In particular, it is observed that the MSE of the MA array in \cite{ma2024sensing} and ``Dense FPA" schemes approaches the CRB only for single target sensing, while the proposed scheme achieves significantly high AoA estimation accuracy even for the scenarios with seven targets.

\begin{figure*}[t]
	\centering
	\subfloat[Proposed MA \label{fig:CorrelProposedMA}]{\includegraphics[width= 4.2 cm]{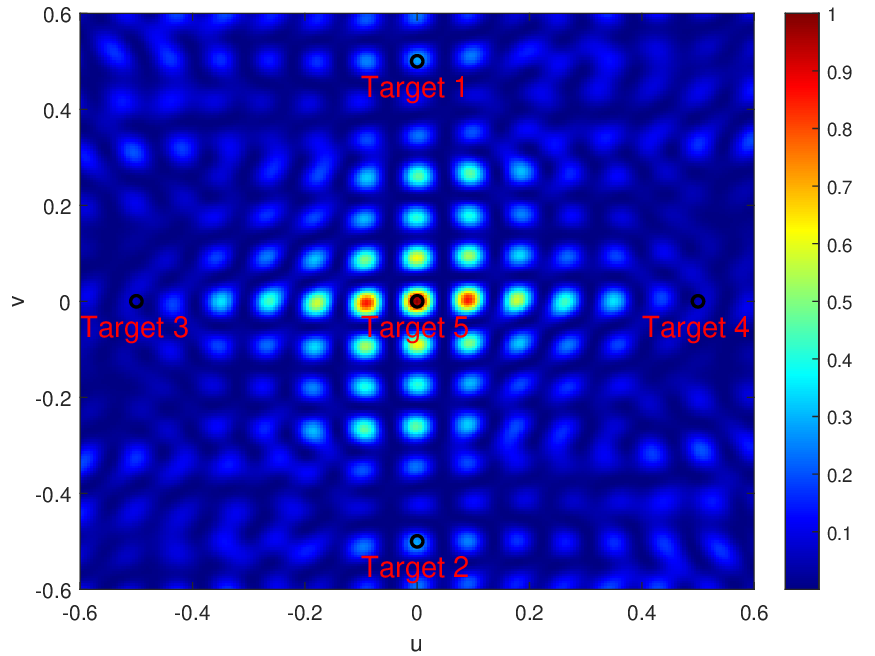}}
	\ \
	\subfloat[MA array in \cite{ma2024sensing} \label{fig:CorrelMAOneTarget}]{\includegraphics[width= 4.2 cm]{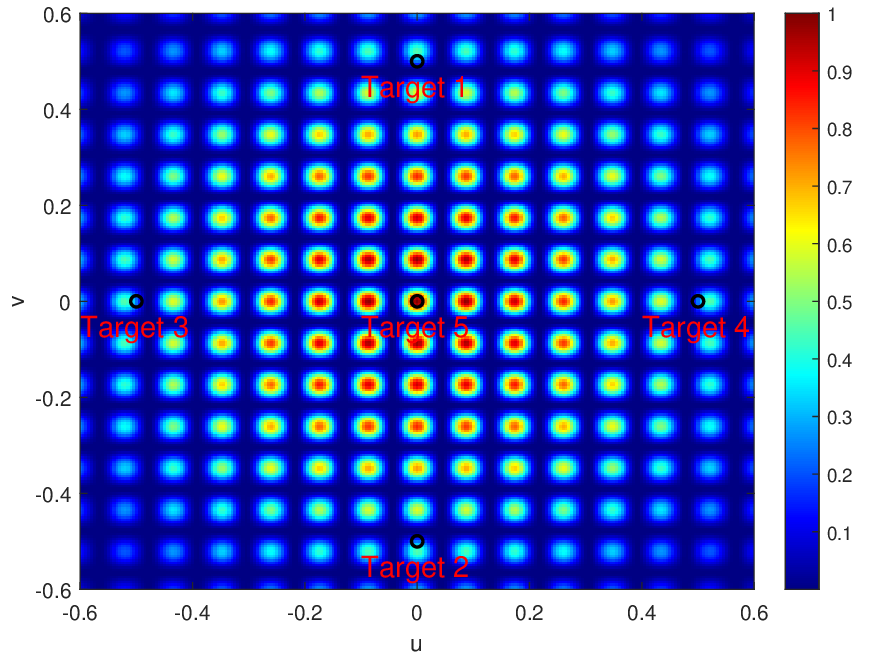}}
	\ \
	\subfloat[Dense UPA\label{fig:CorrelDenseUPA}]{\includegraphics[width= 4.2 cm]{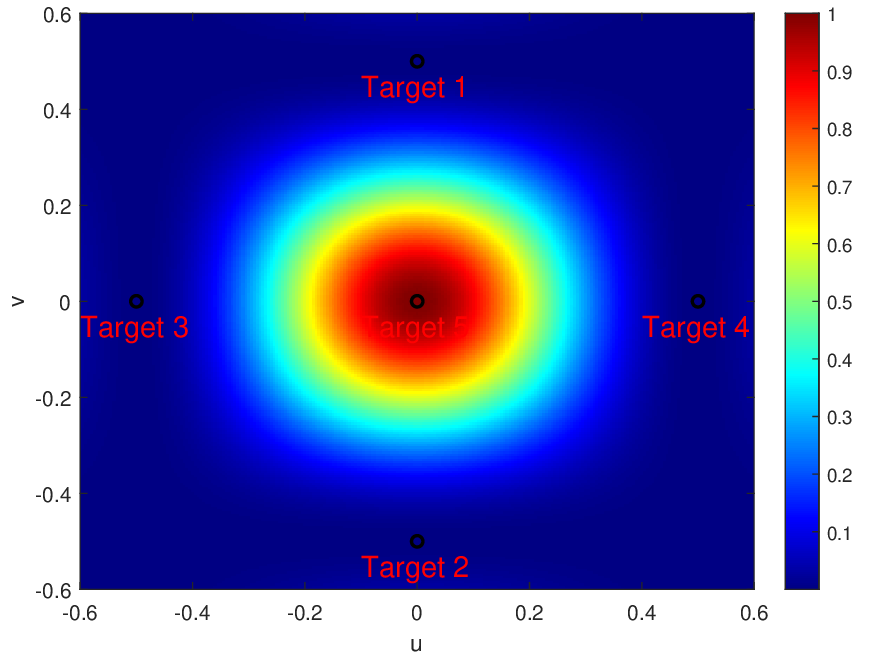}}
	\ \
	\subfloat[Sparse UPA\label{fig:CorrelSparseUPA}]{\includegraphics[width= 4.2 cm]{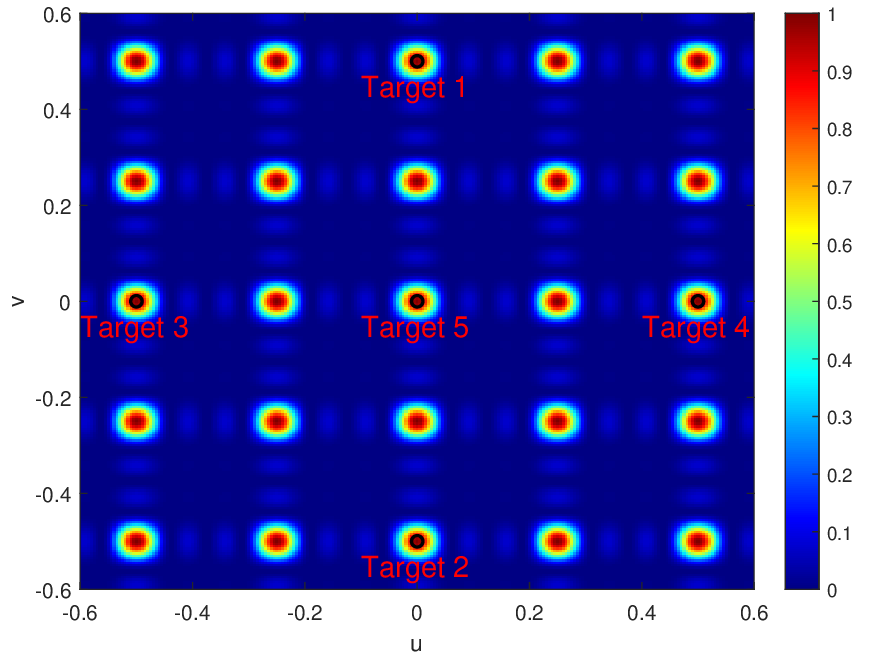}}%
	\caption{Comparison of steering vector correlation with different antennas' positions.}
	\label{fig:Correl}
	\vspace{-2em}
\end{figure*}

In Fig. \ref{fig:RegionSize}, we compare the performance of different schemes versus normalized size of antenna moving region, i.e., $A/\lambda$. It is demonstrated that the CRB performance of all schemes, with the exception of the conventional ``Dense UPA'' method, exhibits an enhancement in the event of an increase in the normalized size of the antenna moving region due to the increase of the achievable array aperture. Notably, the CRB performance gap between the proposed design and the lower-bound becomes smaller with larger antenna moving region sizes by fully exploiting the spatial DoFs via antenna position refinement. In contrast, the CRB performance gap between the proposed design and the MA array in \cite{ma2024sensing} scheme becomes larger with larger antenna moving region sizes. This is because the MA array in \cite{ma2024sensing} induces more side and/or grating lobes at larger apertures, as will be detailed later in Fig. \ref{fig:Correl}\subref{fig:CorrelMAOneTarget}, thereby exacerbating the interference for multi-target sensing. This fact aligns with the observation that the MSE of the MA array in \cite{ma2024sensing} for multi-target AoA estimation decreases and then increases as the normalized size of antenna moving region increases.

%
%

In Fig. \ref{fig:AngleRange}, we compare the performance of different schemes versus angular range of target distribution, i.e., $u_{\max}$ and $v_{\max}$. We can observe that our proposed MA-based design consistently yields substantial gains over all benchmarks, with especially pronounced improvements when multiple targets are densely distributed. This highlights its adaptability to diverse target distributions via effective interference mitigation for multi-target sensing. In addition, it is noted that the CRB performance of ``Sparse UPA'' scheme decreases as the angular range of target distribution increases due to the larger numbers of grating lobes in a given specific local region, which further intensifies the interference between multi-target spatial AoA estimation.

In Fig. \ref{fig:AntennaNumber}, we compare the performance of different schemes versus number of antennas. As can be observed, our proposed scheme always outperforms the benchmark schemes and performs closely to the CRB lower-bound for different numbers of antennas. Furthermore, our proposed scheme demonstrates a significant MSE reduction over all the benchmark schemes with a small number of antennas, e.g., $N =12,16,20$. These results underscore the practical significance of our MA-enhanced multi-target sensing system design, particularly under the constraint of a limited number of antennas.

\begin{figure}[t]
	\centering
	\includegraphics[width= 7 cm]{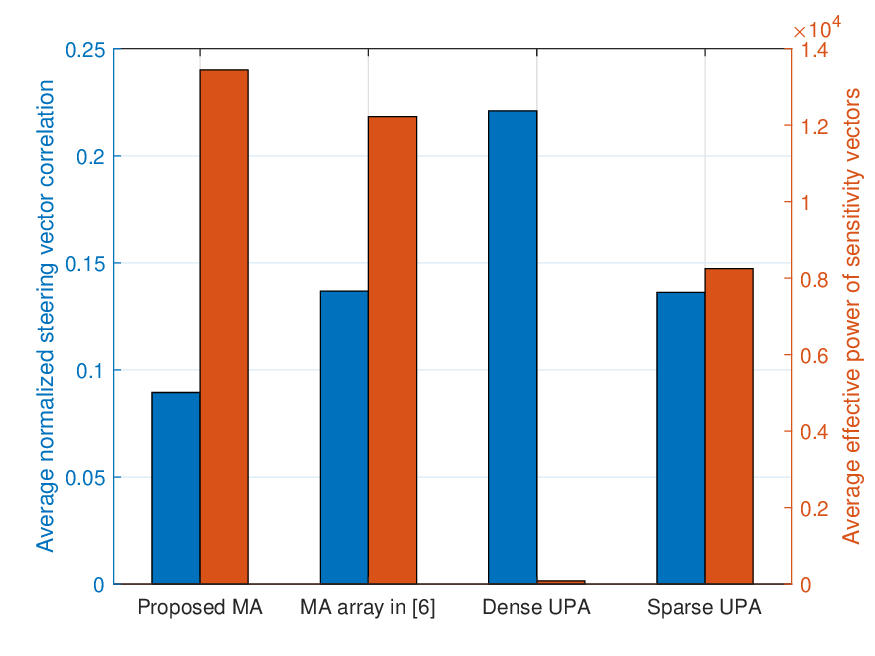}
	\caption{Comparison of coupling and sensitivity for estimating multi-target AoAs.}
	\label{fig:Insights}
	\vspace{-2em}
\end{figure}

\begin{figure*}[t]
	\centering
	\subfloat[Proposed MA, MSE: $3.2 \times 10^{-7}$ \label{fig:MUSICProposedMA}]{\includegraphics[width= 4.2 cm]{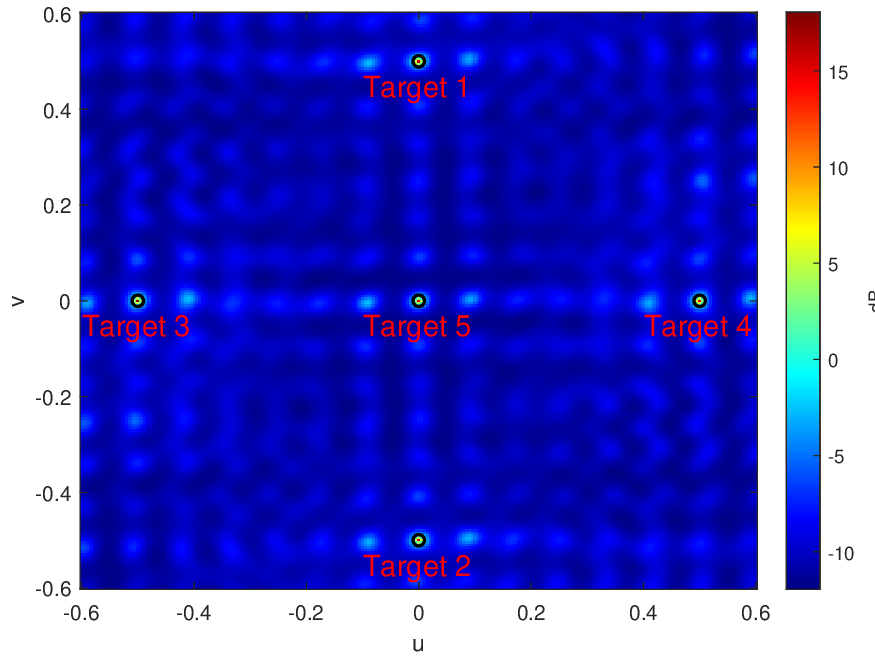}}%
	\ \
	\subfloat[MA array in \cite{ma2024sensing}, MSE: $0.6873$ \label{fig:MUSICMAOneTarget}]{\includegraphics[width= 4.2 cm]{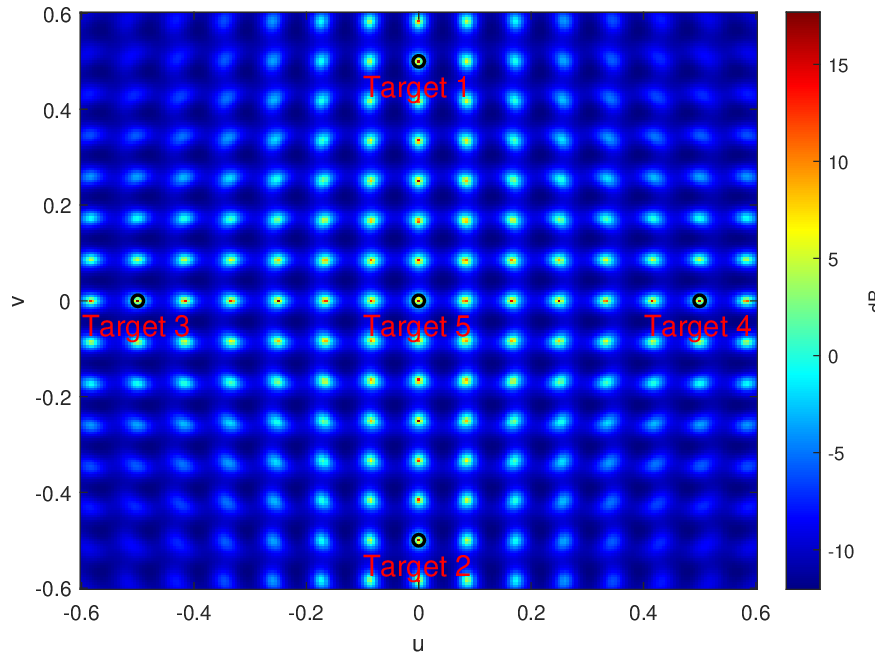}}%
	\ \
	\subfloat[Dense UPA, MSE: $6.78 \times 10^{-5}$\label{fig:MUSICDenseUPA}]{\includegraphics[width= 4.2 cm]{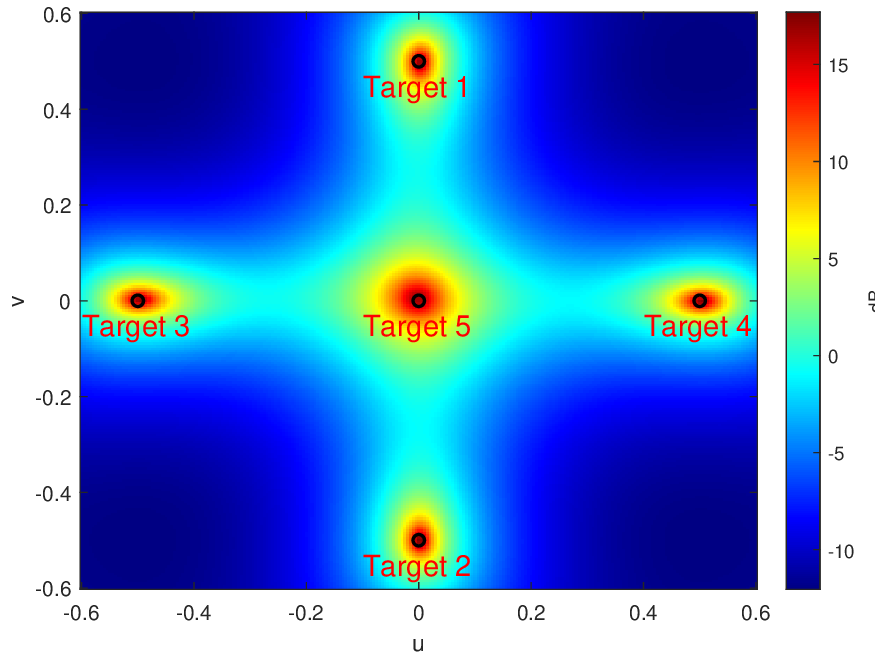}}%
	\ \
	\subfloat[Sparse UPA, MSE: $1.125$\label{fig:MUSICSparseUPA}]{\includegraphics[width= 4.2 cm]{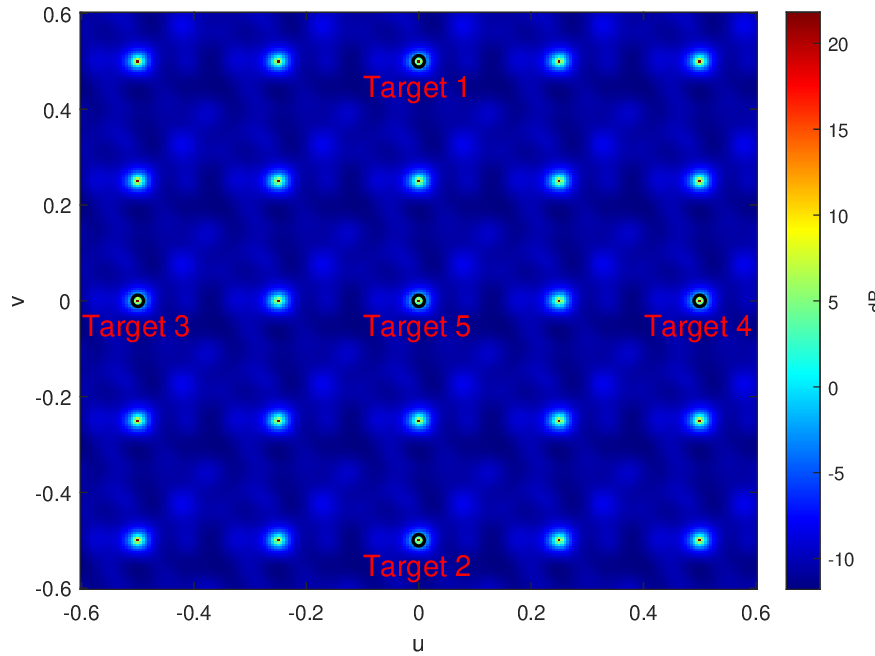}}%
	\caption{Comparison of spatial spectrum under the MUSIC algorithm with different antennas' positions.}
	\label{fig:MUSIC}
	\vspace{-1em}
\end{figure*}

To gain more insights, in Fig. \ref{fig:Insights}, we evaluate the average normalized sensitivity vector correlation and the average effective power of sensitivity vectors for different schemes in multi-target sensing, which are defined as $\rho = \frac{1}{4K(K-1)}\sum\nolimits_{\iota\in \left\{ \rm u,v \right\}}\sum\nolimits_{\iota'\in \left\{ \rm u,v \right\}} \sum\nolimits_{k=1}^{K} \sum\nolimits_{k=1,k\neq k'}^{K}\rho_{k,k'}(\iota,\iota')$ and $\omega = \frac{1}{2K}\sum\nolimits_{\iota\in \left\{ \rm u,v \right\}} \sum\nolimits_{k=1}^{K}\omega_{\iota,k}$, respectively. We can observe that our proposed MA-based design not only reduces the normalized correlation between the sensitivity vectors for different target AoA estimation but also increases the effective power of sensitivity vectors for estimating each target's 2D AoAs, thus achieving a superior CRB performance over the benchmark schemes. This aligns with the theoretical analysis presented in Theorem \ref{theorem}. Furthermore, we present in Fig. \ref{fig:Correl} the normalized steering vector correlation defined as $\frac{1}{N^2} | {\bf{a}}\left( \tilde{{\bf q}},{{\bf{r}}_5} \right)^{\mathsf H}{\bf{a}}\left( \tilde{{\bf q}},{\bar{\bf{r}}} \right)|^2$, for different schemes versus ${\bar{\bf{r}}} \in [-0.6,0.6] \times [-0.6,0.6]$ with the spatial position of target 5 given by ${{\bf{r}}_5} = [0,0]^{\mathsf T}$. It is observed that our proposed MA-based design yields a narrower main lobe and fewer side or grating lobes, thereby enhancing the MSE performance over the benchmark schemes. In addition, the MA array in \cite{ma2024sensing} and the ``Sparse UPA'' scheme suffer from large numbers of side and/or grating lobes, which introduce spurious spatial spectrum peaks in undesired directions as shown in Fig. \ref{fig:MUSIC}, thereby significantly degrading multi-target angle estimation accuracy.


\section{Conclusion} \label{section:s5}
In this paper, we presented a novel wireless sensing system that leverages MA arrays to enhance multi-target sensing performance via antenna position optimization. Specifically, we first characterized the CRB matrix for multi-target AoA estimation as a function of the antenna's positions in the MA array. Then, we formulated an optimization problem to minimize the expectation of the trace of the CRB matrix over random target angles subject to a given distribution via antenna position configuration. To tackle the resultant highly non-convex problem, we employed the Monte Carlo method to address the intractable objective function, and subsequently developed a swarm-based gradient descent algorithm to solve the approximated problem. Additionally, a lower-bound on the sum of CRBs for multi-target AoA estimation was derived. Finally, extensive simulation results were provided to validate the effectiveness and superiority of our proposed MA-based design compared to conventional sensing systems with FPA arrays and single-target-oriented MA arrays, in terms of decreasing both CRB and the actual MSE. Fundamentally, the optimized MA array geometry yields low correlation and high effective power of sensitivity vectors for multi-target AoA estimation, leading to significant CRB performance improvement. The resultant low correlation of steering vectors over multiple targets' directions in the angular domain further helps mitigate angle estimation ambiguity and thereby enhances MSE performance. These results highlight the practical significance of our design for high-accuracy multi-target sensing.

\appendices
\section{Derivation of ${\rm{CRB}}(\tilde{{\bf q}})$} \label{app:A}
First, for ease of notation, we denote
\begin{equation}\small
{{\bf{D}}_{\bm{\omega }}} = \left[ {{{\bf{D}}_{\rm{u}}},{{\bf{D}}_{\rm{v}}}} \right],
\end{equation}
with
\begin{equation}\small
\begin{aligned}
{{\bf{D}}_{\rm{u}}} &= \left[ {\frac{{\partial {\bm{\mu }}}}{{\partial {u_1}}}, \cdots ,\frac{{\partial {\bm{\mu }}}}{{\partial {u_K}}}} \right] \\
&= ({{\bf{S}}^{\mathsf T}} \otimes {{\bf{I}}_N})\left[ {{\bf{e}}_1} \otimes {\dot{\bf{a}}}_{\rm u}\left( {\tilde{{\bf q}},{{\bf{r}}_1}} \right), \cdots ,{{\bf{e}}_K} \otimes {\dot{\bf{a}}}_{\rm u}\left( {\tilde{{\bf q}},{{\bf{r}}_K}} \right) \right]\\
& = ({{\bf{S}}^{\mathsf T}} \otimes {{\bf{I}}_N}) \mathrm{blkdiag}({\dot{\bf{a}}}_{\rm u}\left( {\tilde{{\bf q}},{{\bf{r}}_1}} \right), \cdots, {\dot{\bf{a}}}_{\rm u}\left( {\tilde{{\bf q}},{{\bf{r}}_K}} \right))\\
&  \buildrel \Delta \over= ({{\bf{S}}^{\mathsf T}} \otimes {{\bf{I}}_N}) \widetilde{\bf D}_{\rm u},
\end{aligned}
\end{equation}
\begin{equation}\small
\begin{aligned}
{{\bf{D}}_{\rm{v}}} 
& = ({{\bf{S}}^{\mathsf T}} \otimes {{\bf{I}}_N}) \mathrm{blkdiag}({\dot{\bf{a}}}_{\rm v}\left( {\tilde{{\bf q}},{{\bf{r}}_1}} \right), \cdots, {\dot{\bf{a}}}_{\rm v}\left( {\tilde{{\bf q}},{{\bf{r}}_K}} \right))\\
&  \buildrel \Delta \over= ({{\bf{S}}^{\mathsf T}} \otimes {{\bf{I}}_N}) \widetilde{\bf D}_{\rm v}.
\end{aligned}
\end{equation}
Therefore, we have
\begin{equation}\small
{\bf{D}}_{\bm{\omega }}^{\mathsf H}{{\bf{D}}_{\bm{\omega }}} = {\left[ {{{\bf{D}}_{\rm{u}}},{{\bf{D}}_{\rm{v}}}} \right]^{\mathsf H}}\left[ {{{\bf{D}}_{\rm{u}}},{{\bf{D}}_{\rm{v}}}} \right] = \left[ {\begin{array}{*{20}{c}}
	{{\bf{D}}_{\rm{u}}^{\mathsf H}{{\bf{D}}_{\rm{u}}}}&{{\bf{D}}_{\rm{u}}^{\mathsf H}{{\bf{D}}_{\rm{v}}}}\\
	{{\bf{D}}_{\rm{v}}^{\mathsf H}{{\bf{D}}_{\rm{u}}}}&{{\bf{D}}_{\rm{v}}^{\mathsf H}{{\bf{D}}_{\rm{v}}}}
	\end{array}} \right],
\end{equation}
where
\begin{equation}\small
\begin{aligned}
{\bf{D}}_{\rm{u}}^{\mathsf H}{{\bf{D}}_{\rm{u}}}& = \widetilde{\bf D}_{\rm u}^{\mathsf H}({{\bf{S}}^*} \otimes {{\bf{I}}_N})({{\bf{S}}^{\mathsf T}} \otimes {{\bf{I}}_N})\widetilde{\bf D}_{\rm u} \\
&= \widetilde{\bf D}_{\rm u}^{\mathsf H}\left( {{{\bf{S}}^*}{{\bf{S}}^{\mathsf T}} \otimes {{\bf{I}}_N}} \right)\widetilde{\bf D}_{\rm u}= {{\bf R}_{\bf{S}}^{\mathsf T}} \odot \dot{\bf{ A}}_{\rm{u}}^{\mathsf H}{\dot{\bf{ A}}_{\rm{u}}},
\end{aligned}
\end{equation}
\begin{equation}\small
\begin{aligned}
{\bf{D}}_{\rm{u}}^{\mathsf H}{{\bf{D}}_{\rm{v}}} = {{\bf R}_{\bf{S}}^{\mathsf T}} \odot \dot{\bf{ A}}_{\rm{u}}^{\mathsf H}{\dot{\bf{ A}}_{\rm{v}}},
\end{aligned}
\end{equation}
\begin{equation}\small
\begin{aligned}
{\bf{D}}_{\rm{v}}^{\mathsf H}{{\bf{D}}_{\rm{u}}} = {{\bf R}_{\bf{S}}^{\mathsf T}} \odot \dot{\bf{ A}}_{\rm{v}}^{\mathsf H}{\dot{\bf{ A}}_{\rm{u}}},
\end{aligned}
\end{equation}
\begin{equation}\small
\begin{aligned}
{\bf{D}}_{\rm{v}}^{\mathsf H}{{\bf{D}}_{\rm{v}}} = {{\bf R}_{\bf{S}}^{\mathsf T}} \odot \dot{\bf{ A}}_{\rm{v}}^{\mathsf H}{\dot{\bf{ A}}_{\rm{v}}}.
\end{aligned}
\end{equation}

Similarly, we denote ${{\bf{D}}_{\bm{\zeta }}} = \left[ {{{\bf{D}}_{\rm{R}}},{{\bf{D}}_{\rm{I}}}} \right]$ with
\begin{equation}\small
\begin{aligned}
{{\bf{D}}_{\rm{R}}}& = \left[ {\frac{{\partial {\bm{\mu }}}}{{\partial \Re \{ {{\bf{s}}_1}\} }}, \cdots ,\frac{{\partial {\bm{\mu }}}}{{\partial \Re \{ {{\bf{s}}_K}\} }}} \right]\\
& = \left[ {{{\bf{I}}_T} \otimes {\bf{a}}\left( {\tilde{\bf{q}},{{\bf{r}}_1}} \right), \cdots ,{{\bf{I}}_T} \otimes {\bf{a}}\left( {\tilde{\bf{q}},{{\bf{r}}_K}} \right)} \right],
\end{aligned}
\end{equation}
\begin{equation}\small
\begin{aligned}
{{\bf{D}}_{\rm{I}}}& = \left[ {\frac{{\partial {\bm{\mu }}}}{{\partial \Im \{ {{\bf{s}}_1}\} }}, \cdots ,\frac{{\partial {\bm{\mu }}}}{{\partial \Im \{ {{\bf{s}}_K}\} }}} \right]\\
& = {\rm j}\left[ {{{\bf{I}}_T} \otimes {\bf{a}}\left( {\tilde{\bf{q}},{{\bf{r}}_1}} \right), \cdots ,{{\bf{I}}_T} \otimes {\bf{a}}\left( {\tilde{\bf{q}},{{\bf{r}}_K}} \right)} \right].
\end{aligned}
\end{equation}
Then, we have
\begin{equation}\small
\begin{aligned}
&{\bf{D}}_{\bm{\omega }}^{\mathsf H}{{\bf{D}}_{\bm{\zeta }}} = {\left[ {({{\bf{S}}^{\mathsf T}} \otimes {{\bf{I}}_N})\widetilde{\bf D}_{\rm u},({{\bf{S}}^{\mathsf T}} \otimes {{\bf{I}}_N})\widetilde{\bf D}_{\rm v}} \right]^{\mathsf H}}\left[ {{{\bf{D}}_{\rm{R}}},{\rm{j}}{{\bf{D}}_{\rm{R}}}} \right]\\
&= \left[ {\begin{array}{*{20}{c}}
	{\widetilde{\bf D}_{\rm u}^{\mathsf H}({{\bf{S}}^*} \otimes {{\bf{I}}_N}){{\bf{D}}_{\rm{R}}}}&{{\rm{j}}\widetilde{\bf D}_{\rm u}^{\mathsf H}({{\bf{S}}^*} \otimes {{\bf{I}}_N}){{\bf{D}}_{\rm{R}}}}\\
	\widetilde{\bf D}_{\rm v}^{\mathsf H}({{\bf{S}}^*} \otimes {{\bf{I}}_N}){{\bf{D}}_{\rm{R}}}&{\rm{j}}\widetilde{\bf D}_{\rm v}^{\mathsf H}({{\bf{S}}^*} \otimes {{\bf{I}}_N}){{\bf{D}}_{\rm{R}}}
	\end{array}} \right],
\end{aligned}
\end{equation}
\begin{equation}\small
\begin{aligned}
&{\bf{D}}_{\bm{\zeta }}^{\mathsf H}{{\bf{D}}_{\bm{\omega }}} = {({\bf{D}}_{\bm{\omega }}^{\mathsf H}{{\bf{D}}_{\bm{\zeta }}})^{\mathsf H}}\\
&= \left[ {\begin{array}{*{20}{c}}
	{{{\bf{D}}_{\rm{R}}^{\mathsf H}}({{\bf{S}}^{\mathsf T}} \otimes {{\bf{I}}_N})\widetilde{\bf D}_{\rm u}}&{{{\bf{D}}_{\rm{R}}^{\mathsf H}}({{\bf{S}}^{\mathsf T}} \otimes {{\bf{I}}_N})\widetilde{\bf D}_{\rm v}}\\
	{-\rm{j}}{{\bf{D}}_{\rm{R}}^{\mathsf H}}({{\bf{S}}^{\mathsf T}} \otimes {{\bf{I}}_N})\widetilde{\bf D}_{\rm u}&{-\rm{j}}{{\bf{D}}_{\rm{R}}^{\mathsf H}}({{\bf{S}}^{\mathsf T}} \otimes {{\bf{I}}_N})\widetilde{\bf D}_{\rm v}
	\end{array}} \right],
\end{aligned}
\end{equation}
and
\begin{equation}\small
\begin{aligned}
&{\bf{D}}_{\bm{\zeta }}^{\mathsf H}{{\bf{D}}_{\bm{\zeta }}} = {\left[ {{{\bf{D}}_{\rm{R}}},{\rm{j}}{{\bf{D}}_{\rm{R}}}} \right]^{\mathsf H}}\left[ {{{\bf{D}}_{\rm{R}}},{\rm{j}}{{\bf{D}}_{\rm{R}}}} \right] = \left[ {\begin{array}{*{20}{c}}
	{{\bf{D}}_{\rm{R}}^{\mathsf H}{{\bf{D}}_{\rm{R}}}}&{{\rm{j}}{\bf{D}}_{\rm{R}}^{\mathsf H}{{\bf{D}}_{\rm{R}}}}\\
	{{\rm{ - j}}{\bf{D}}_{\rm{R}}^{\mathsf H}{{\bf{D}}_{\rm{R}}}}&{{\bf{D}}_{\rm{R}}^{\mathsf H}{{\bf{D}}_{\rm{R}}}}
	\end{array}} \right],
\end{aligned}
\end{equation}
with
\begin{equation}\small
{{\bf{D}}_{\rm{R}}^{\mathsf H}{{\bf{D}}_{\rm{R}}}} = {{{\bf{A}}^{\mathsf H}}{\bf{A}} \otimes {{\bf{I}}_T}}.
\end{equation}

Then, ${\bf F}_{22}$ can be further written as
\begin{equation}\small
\begin{aligned}
&{{\bf{F}}_{22}} = \Re \left\{ {\bf{D}}_{\bm{\zeta }}^{\mathsf H}{{\bf{D}}_{\bm{\zeta }}} \right\} = \left[ {\begin{array}{*{20}{c}}
	{\Re \left\{ {{{\bf{A}}^{\mathsf H}}{\bf{A}} \otimes {{\bf{I}}_T}} \right\}}&{ - \Im \left\{ {{{\bf{A}}^{\mathsf H}}{\bf{A}} \otimes {{\bf{I}}_T}} \right\}}\\
	{\Im \left\{ {{{\bf{A}}^{\mathsf H}}{\bf{A}} \otimes {{\bf{I}}_T}} \right\}}&{\Re \left\{ {{{\bf{A}}^{\mathsf H}}{\bf{A}} \otimes {{\bf{I}}_T}} \right\}}
	\end{array}} \right].
\end{aligned}
\end{equation}
It is noted that for ${\bf{\Xi }} =  {{{\bf{A}}^{\mathsf H}}{\bf{A}} \otimes {{\bf{I}}_T}} \in \mathbb{C}^{KT \times KT}$, we have
\begin{equation}\small
\Re \left\{ {\bf{\Xi }} \right\}\Re \left\{ {{{\bf{\Xi }}^{ - 1}}} \right\} - \Im \left\{ {\bf{\Xi }} \right\}\Im \left\{ {{{\bf{\Xi }}^{ - 1}}} \right\} = \Re \left\{ {{\bf{\Xi }}{{\bf{\Xi }}^{ - 1}}} \right\} = {{\bf{I}}_{KT}},
\end{equation}
\begin{equation}\small
\Re \left\{ {\bf{\Xi }} \right\} {  \Im \left\{ {{{\bf{\Xi }}^{ - 1}}} \right\}}  +  { \Im \left\{ {\bf{\Xi }} \right\}} \Re \left\{ {{{\bf{\Xi }}^{ - 1}}} \right\} =  \Im \left\{ {{\bf{\Xi }}{{\bf{\Xi }}^{ - 1}}} \right\} = {{\bf{0}}_{KT}}.
\end{equation}
Therefore, the inverse of ${{\bf{F}}_{22}}$ can be given by
\begin{equation}\small
{\bf{F}}_{22}^{ - 1} = \left[ {\begin{array}{*{20}{c}}
{\Re \left\{ {{{\left( {{{\bf{A}}^{\mathsf H}}{\bf{A}}} \right)}^{ - 1}} \otimes {{\bf{I}}_T}} \right\}}&{ - \Im \left\{ {{{\left( {{{\bf{A}}^{\mathsf H}}{\bf{A}}} \right)}^{ - 1}} \otimes {{\bf{I}}_T}} \right\}}\\
{\Im \left\{ {{{\left( {{{\bf{A}}^{\mathsf H}}{\bf{A}}} \right)}^{ - 1}} \otimes {{\bf{I}}_T}} \right\}}&{\Re \left\{ {{{\left( {{{\bf{A}}^{\mathsf H}}{\bf{A}}} \right)}^{ - 1}} \otimes {{\bf{I}}_T}} \right\}}
\end{array}} \right].
\end{equation}

Next, we can derive ${{\bf{F}}_{12}}{\bf{F}}_{22}^{ - 1}{{\bf{F}}_{21}}$ as (\ref{inverse}) shown at the top of the next page.
\begin{figure*}[hbt]
	\begin{equation} \label{inverse} \small
	\begin{split}
	&{{\bf{F}}_{12}}{\bf{F}}_{22}^{ - 1}{{\bf{F}}_{21}}= \Re\left\{{\bf{D}}_{\bm{\omega }}^{\mathsf H}{{\bf{D}}_{\bm{\zeta }}}\right\}{\bf{F}}_{22}^{ - 1}\Re\left\{{\bf{D}}_{\bm{\zeta }}^{\mathsf H}{{\bf{D}}_{\bm{\omega }}}\right\}\\
	& = \left[ {\begin{array}{*{20}{c}}
		{\Re \left\{ {\widetilde{\bf D}_{\rm u}^{\mathsf H}({{\bf{S}}^*} \otimes {{\bf{I}}_N}){{\bf{D}}_{\rm{R}}}({{({{\bf{A}}^{\mathsf H}}{\bf{A}})}^{ - 1}} \otimes {{\bf{I}}_T}){\bf{D}}_{\rm{R}}^{\mathsf H}({{\bf{S}}^{\mathsf T}} \otimes {{\bf{I}}_N})\widetilde{\bf D}_{\rm u}} \right\}}&{\Re \left\{ {\widetilde{\bf D}_{\rm u}^{\mathsf H}({{\bf{S}}^*} \otimes {{\bf{I}}_N}){{\bf{D}}_{\rm{R}}} ({{({{\bf{A}}^{\mathsf H}}{\bf{A}})}^{ - 1}}\otimes {{\bf{I}}_T}){\bf{D}}_{\rm{R}}^{\mathsf H}({{\bf{S}}^{\mathsf T}} \otimes {{\bf{I}}_N})\widetilde{\bf D}_{\rm v}} \right\}}\\
		{\Re \left\{ {\widetilde{\bf D}_{\rm v}^{\mathsf H}({{\bf{S}}^*} \otimes {{\bf{I}}_N}){{\bf{D}}_{\rm{R}}}({{({{\bf{A}}^H}{\bf{A}})}^{ - 1}} \otimes {{\bf{I}}_T}){\bf{D}}_{\rm{R}}^{\mathsf H}({{\bf{S}}^{\mathsf T}} \otimes {{\bf{I}}_N})\widetilde{\bf D}_{\rm u}} \right\}}&{\Re \left\{ {\widetilde{\bf D}_{\rm v}^{\mathsf H}({{\bf{S}}^*} \otimes {{\bf{I}}_N}){{\bf{D}}_{\rm{R}}}({{({{\bf{A}}^H}{\bf{A}})}^{ - 1}} \otimes {{\bf{I}}_T}) {\bf{D}}_{\rm{R}}^{\mathsf H}({{\bf{S}}^{\mathsf T}} \otimes {{\bf{I}}_N})\widetilde{\bf D}_{\rm v}} \right\}}
		\end{array}} \right]\\
	& = \left[ {\begin{array}{*{20}{c}}
		{\Re \left\{ {\widetilde{\bf D}_{\rm u}^{\mathsf H}({{\bf{S}}^*} \otimes {{\bf{I}}_N}){({{\bf{I}}_T} \otimes {\bf{A}}{{({{\bf{A}}^{\mathsf H}}{\bf{A}})}^{ - 1}}{{\bf{A}}^{\mathsf H}})}({{\bf{S}}^{\mathsf T}} \otimes {{\bf{I}}_N})\widetilde{\bf D}_{\rm u}} \right\}}&{\Re \left\{ {\widetilde{\bf D}_{\rm u}^{\mathsf H}({{\bf{S}}^*} \otimes {{\bf{I}}_N}){({{\bf{I}}_T} \otimes {\bf{A}}{{({{\bf{A}}^{\mathsf H}}{\bf{A}})}^{ - 1}}{{\bf{A}}^{\mathsf H}})}({{\bf{S}}^{\mathsf T}} \otimes {{\bf{I}}_N})\widetilde{\bf D}_{\rm v}} \right\}}\\
		{\Re \left\{ {\widetilde{\bf D}_{\rm v}^{\mathsf H}({{\bf{S}}^*} \otimes {{\bf{I}}_N}) {({{\bf{I}}_T} \otimes {\bf{A}}{{({{\bf{A}}^{\mathsf H}}{\bf{A}})}^{ - 1}}{{\bf{A}}^{\mathsf H}})} ({{\bf{S}}^{\mathsf T}} \otimes {{\bf{I}}_N})\widetilde{\bf D}_{\rm u}} \right\}}&{\Re \left\{ {\widetilde{\bf D}_{\rm v}^{\mathsf H}({{\bf{S}}^*} \otimes {{\bf{I}}_N}){({{\bf{I}}_T} \otimes {\bf{A}}{{({{\bf{A}}^{\mathsf H}}{\bf{A}})}^{ - 1}}{{\bf{A}}^{\mathsf H}})}({{\bf{S}}^{\mathsf T}} \otimes {{\bf{I}}_N})\widetilde{\bf D}_{\rm v}} \right\}}
		\end{array}} \right]\\
	&=\left[ {\begin{array}{*{20}{c}}
		{\Re \left\{ {\widetilde{\bf D}_{\rm u}^{\mathsf H} {({{\bf{S}}^*}{{\bf{S}}^{\mathsf T}} \otimes {\bf{A}}{{({{\bf{A}}^{\mathsf H}}{\bf{A}})}^{ - 1}}{{\bf{A}}^{\mathsf H}})} \widetilde{\bf D}_{\rm u}} \right\}}&{\Re \left\{ {\widetilde{\bf D}_{\rm u}^{\mathsf H}{({{\bf{S}}^*}{{\bf{S}}^{\mathsf T}} \otimes {\bf{A}}{{({{\bf{A}}^{\mathsf H}}{\bf{A}})}^{ - 1}}{{\bf{A}}^{\mathsf H}})}\widetilde{\bf D}_{\rm v}} \right\}}\\
		{\Re \left\{ {\widetilde{\bf D}_{\rm v}^{\mathsf H}{({{\bf{S}}^*}{{\bf{S}}^{\mathsf T}} \otimes {\bf{A}}{{({{\bf{A}}^{\mathsf H}}{\bf{A}})}^{ - 1}}{{\bf{A}}^{\mathsf H}})}\widetilde{\bf D}_{\rm u}} \right\}}&{\Re \left\{ {\widetilde{\bf D}_{\rm v}^{\mathsf H}{({{\bf{S}}^*}{{\bf{S}}^{\mathsf T}} \otimes {\bf{A}}{{({{\bf{A}}^{\mathsf H}}{\bf{A}})}^{ - 1}}{{\bf{A}}^{\mathsf H}})}\widetilde{\bf D}_{\rm v}} \right\}}
		\end{array}} \right]\\
	&=\Re \left\{ {\left[ {\begin{array}{*{20}{c}}
			{{\bf{R}}_{\bf{S}}^{\mathsf T} \odot \dot{\bf{ A}}_{\rm{u}}^{\mathsf H}{\bf{A}}{{({{\bf{A}}^{\mathsf H}}{\bf{A}})}^{ - 1}}{{\bf{A}}^{\mathsf H}}{{\dot{\bf{ A}}}_{\rm{u}}}}&{{\bf{R}}_{\bf{S}}^{\mathsf T} \odot \dot{\bf{ A}}_{\rm{u}}^{\mathsf H}{\bf{A}}{{({{\bf{A}}^{\mathsf H}}{\bf{A}})}^{ - 1}}{{\bf{A}}^{\mathsf H}}{{\dot{\bf{ A}}}_{\rm{v}}}}\\
			{{\bf{R}}_{\bf{S}}^{\mathsf T} \odot \dot{\bf{ A}}_{\rm{v}}^{\mathsf H}{\bf{A}}{{({{\bf{A}}^{\mathsf H}}{\bf{A}})}^{ - 1}}{{\bf{A}}^{\mathsf H}}{{\dot{\bf{ A}}}_{\rm{u}}}}&{{\bf{R}}_{\bf{S}}^{\mathsf T} \odot \dot{\bf{ A}}_{\rm{v}}^{\mathsf H}{\bf{A}}{{({{\bf{A}}^{\mathsf H}}{\bf{A}})}^{ - 1}}{{\bf{A}}^{\mathsf H}}{{\dot{\bf{ A}}}_{\rm{v}}}}
			\end{array}} \right]} \right\}
	\end{split}
	\end{equation}
	\vspace*{-1em} 
	\hrulefill
	\vspace*{-0.5em}
\end{figure*}
Then, we can obtain the CRB matrix for multi-target 2D spatial AoA estimation by employing the Schur complement theorem as (\ref{crb0}) shown at the top of the next page.
\begin{figure*}[htb]
\begin{equation} \label{crb0} \small
\begin{aligned}
&{\rm{CRB}}(\tilde{{\bf q}}) = \frac{{{\sigma ^2}}}{2}{\left( {{{\bf{F}}_{11}} - {{\bf{F}}_{12}}{\bf{F}}_{22}^{ - 1}{{\bf{F}}_{21}}} \right)^{ - 1}}\\
& = \frac{{{\sigma ^2}}}{2} \left( \Re \left\{\left[ {\begin{array}{*{20}{c}}
	{  {{\bf{R}}_{\bf{S}}^{\mathsf T} \odot \dot{\bf{ A}}_{\rm{u}}^{\mathsf H}{{\dot{\bf{ A}}}_{\rm{u}}}} }&{ {{\bf{R}}_{\bf{S}}^{\mathsf T} \odot \dot{\bf{ A}}_{\rm{u}}^{\mathsf H}{{\dot{\bf{ A}}}_{\rm{v}}}} }\\
	{  {{\bf{R}}_{\bf{S}}^{\mathsf T} \odot \dot{\bf{ A}}_{\rm{v}}^{\mathsf H}{{\dot{\bf{ A}}}_{\rm{u}}}} }&{ {{\bf{R}}_{\bf{S}}^{\mathsf T} \odot \dot{\bf{ A}}_{\rm{v}}^{\mathsf H}{{\dot{\bf{ A}}}_{\rm{v}}}}}
	\end{array}} \right]\right\} - \Re \left\{ {\left[ {\begin{array}{*{20}{c}}
		{{\bf{R}}_{\bf{S}}^{\mathsf T} \odot \dot{\bf{ A}}_{\rm{u}}^{\mathsf H}{\bf{A}}{{({{\bf{A}}^{\mathsf H}}{\bf{A}})}^{ - 1}}{{\bf{A}}^{\mathsf H}}{{\dot{\bf{ A}}}_{\rm{u}}}}&{{\bf{R}}_{\bf{S}}^{\mathsf T} \odot \dot{\bf{ A}}_{\rm{u}}^{\mathsf H}{\bf{A}}{{({{\bf{A}}^{\mathsf H}}{\bf{A}})}^{ - 1}}{{\bf{A}}^{\mathsf H}}{{\dot{\bf{ A}}}_{\rm{v}}}}\\
		{{\bf{R}}_{\bf{S}}^{\mathsf T} \odot \dot{\bf{ A}}_{\rm{v}}^{\mathsf H}{\bf{A}}{{({{\bf{A}}^{\mathsf H}}{\bf{A}})}^{ - 1}}{{\bf{A}}^{\mathsf H}}{{\dot{\bf{ A}}}_{\rm{u}}}}&{{\bf{R}}_{\bf{S}}^{\mathsf T} \odot \dot{\bf{ A}}_{\rm{v}}^{\mathsf H}{\bf{A}}{{({{\bf{A}}^{\mathsf H}}{\bf{A}})}^{ - 1}}{{\bf{A}}^{\mathsf H}}{{\dot{\bf{ A}}}_{\rm{v}}}}
		\end{array}} \right]} \right\} \right)^{-1}\\
&= \frac{{{\sigma ^2}}}{2} \left( \Re \left\{\left[ {\begin{array}{*{20}{c}}
	{  {{\bf{R}}_{\bf{S}}^{\mathsf T} \odot \dot{\bf{ A}}_{\rm{u}}^{\mathsf H} ({{\bf{I}}_N} - {\bf{A}}{{({{\bf{A}}^{\mathsf H}}{\bf{A}})}^{ - 1}}{{\bf{A}}^{\mathsf H}}){{\dot{\bf{ A}}}_{\rm{u}}}} }&{ {{\bf{R}}_{\bf{S}}^{\mathsf T} \odot \dot{\bf{ A}}_{\rm{u}}^{\mathsf H}({{\bf{I}}_N} - {\bf{A}}{{({{\bf{A}}^{\mathsf H}}{\bf{A}})}^{ - 1}}{{\bf{A}}^{\mathsf H}}){{\dot{\bf{ A}}}_{\rm{v}}}} }\\
	{  {{\bf{R}}_{\bf{S}}^{\mathsf T} \odot \dot{\bf{ A}}_{\rm{v}}^{\mathsf H}({{\bf{I}}_N} - {\bf{A}}{{({{\bf{A}}^{\mathsf H}}{\bf{A}})}^{ - 1}}{{\bf{A}}^{\mathsf H}}){{\dot{\bf{ A}}}_{\rm{u}}}} }&{ {{\bf{R}}_{\bf{S}}^{\mathsf T} \odot \dot{\bf{ A}}_{\rm{v}}^{\mathsf H}({{\bf{I}}_N} - {\bf{A}}{{({{\bf{A}}^{\mathsf H}}{\bf{A}})}^{ - 1}}{{\bf{A}}^{\mathsf H}}){{\dot{\bf{ A}}}_{\rm{v}}}}}
	\end{array}} \right]\right\} \right)^{-1}\\
&=\frac{{{\sigma ^2}}}{2}{\left( {\Re \left\{ {({{\bf{1}}_{2}} \otimes {\bf{R}}_{\bf{S}}^{\mathsf T}) \odot {{\dot{\bf{ A}}}^{\mathsf H}}{\bf{\Pi }}_{\bf{A}}^ \bot \dot{\bf{ A}}} \right\}} \right)^{ - 1}}
\end{aligned}
\end{equation}
	\vspace*{-1em} 
\hrulefill
\vspace*{-0.5em}
\end{figure*}

This completes the derivation.
\vspace{-1em}
\section{Proof of Theorem \ref{theorem}} \label{app:B}
We begin by proving that (a) in (\ref{lower}) holds. For ease of notation, we first denote
\begin{equation}\small
{\bf{\Phi }}({{\bf{R}}_{\bf{S}}}) = \frac{2}{{{\sigma ^2}}}\Re \left\{\left[ {\begin{array}{*{20}{c}}
	{  {{\bf{R}}_{\bf{S}}^{\mathsf T} \odot \dot{\bf{ A}}_{\rm{u}}^{\mathsf H} {\bf{\Pi }}_{\bf{A}}^ \bot{{\dot{\bf{ A}}}_{\rm{u}}}} }&{ {{\bf{R}}_{\bf{S}}^{\mathsf T} \odot \dot{\bf{ A}}_{\rm{u}}^{\mathsf H}{\bf{\Pi }}_{\bf{A}}^ \bot{{\dot{\bf{ A}}}_{\rm{v}}}} }\\
	{  {{\bf{R}}_{\bf{S}}^{\mathsf T} \odot \dot{\bf{ A}}_{\rm{v}}^{\mathsf H}{\bf{\Pi }}_{\bf{A}}^ \bot{{\dot{\bf{ A}}}_{\rm{u}}}} }&{ {{\bf{R}}_{\bf{S}}^{\mathsf T} \odot \dot{\bf{ A}}_{\rm{v}}^{\mathsf H}{\bf{\Pi }}_{\bf{A}}^ \bot{{\dot{\bf{ A}}}_{\rm{v}}}}}
	\end{array}} \right]\right\},
\end{equation}
and
\begin{equation}\small
{\bf{\Psi }}({\widetilde{\bf{R}}_{\bf{S}}}) = \frac{2}{{{\sigma ^2}}}\Re \left\{\left[ {\begin{array}{*{20}{c}}
	{  {\widetilde{\bf{R}}_{\bf{S}}^{\mathsf T} \odot \dot{\bf{ A}}_{\rm{u}}^{\mathsf H} {\bf{\Pi }}_{\bf{A}}^ \bot{{\dot{\bf{ A}}}_{\rm{u}}}} }&{ {\widetilde{\bf{R}}_{\bf{S}}^{\mathsf T} \odot \dot{\bf{ A}}_{\rm{u}}^{\mathsf H}{\bf{\Pi }}_{\bf{A}}^ \bot{{\dot{\bf{ A}}}_{\rm{v}}}} }\\
	{  {\widetilde{\bf{R}}_{\bf{S}}^{\mathsf T} \odot \dot{\bf{ A}}_{\rm{v}}^{\mathsf H}{\bf{\Pi }}_{\bf{A}}^ \bot{{\dot{\bf{ A}}}_{\rm{u}}}} }&{ {\widetilde{\bf{R}}_{\bf{S}}^{\mathsf T} \odot \dot{\bf{ A}}_{\rm{v}}^{\mathsf H}{\bf{\Pi }}_{\bf{A}}^ \bot{{\dot{\bf{ A}}}_{\rm{v}}}}}
	\end{array}} \right]\right\},
\end{equation}
where ${\widetilde{\bf{R}}_{\bf{S}}} = {\rm diag}(P_{\rm s},P_{\rm s},\cdots,P_{\rm s})\in \mathbb R^{K\times K}$. Then, we have
\begin{equation}\small
{\bf{\Psi }}({\widetilde{\bf{R}}_{\bf{S}}}) = \sum\limits_{k = 1}^K {{{\bf{P}}_k}{\bf{\Phi }}({{\bf{R}}_{\bf{S}}}){{\bf{P}}_k}},
\end{equation}
where ${{\bf{P}}_k} \in  \mathbb R^{2K\times 2K}$ is given by
\begin{equation}\small
{{\bf{P}}_k} = \left[ {\begin{array}{*{20}{c}}
	{{{\bf{E}}_{k,k}}}&{{{\bf{0}}_{K}}}\\
	{{{\bf{0}}_{K}}}&{{{\bf{E}}_{k,k}}}
	\end{array}} \right],
\end{equation}
with ${{\bf{E}}_{k,k}} \in {\mathbb R^{K \times K}}$ denoting the matrix that the entry in the $k$-th row and $k$-th column is $1$ and all other entries are $0$. Since ${\bf{\Phi }}({{\bf{R}}_{\bf{S}}})$ is positive definite and invertible, i.e., ${\bf{\Phi }}({{\bf{R}}_{\bf{S}}}) \succ 0$, we have
\begin{equation}\small
\left[ {\begin{array}{*{20}{c}}
	{{\bf{\Phi }}({{\bf{R}}_{\bf{S}}})}&{{{\bf{I}}_{2K}}}\\
	{{{\bf{I}}_{2K}}}&{{\bf{\Phi }}{{({{\bf{R}}_{\bf{S}}})}^{ - 1}}}
	\end{array}} \right] \succeq 0,
\end{equation}
which holds according to the Schur complement theorem. Then, we have
\begin{equation}\small
\begin{aligned}
& \left[ {\begin{array}{*{20}{c}}
	{{{\bf{P}}_k}}&{{{\bf{0}}_{2K}}}\\
	{{{\bf{0}}_{2K}}}&{{{\bf{P}}_k}}
	\end{array}} \right]\left[ {\begin{array}{*{20}{c}}
	{{\bf{\Phi }}({{\bf{R}}_{\bf{S}}})}&{{{\bf{I}}_{2K}}}\\
	{{{\bf{I}}_{2K}}}&{{\bf{\Phi }}{{({{\bf{R}}_{\bf{S}}})}^{ - 1}}}
	\end{array}} \right]\left[ {\begin{array}{*{20}{c}}
	{{{\bf{P}}_k}}&{{{\bf{0}}_{2K}}}\\
	{{{\bf{0}}_{2K}}}&{{{\bf{P}}_k}}
	\end{array}} \right]\\
&=\left[ {\begin{array}{*{20}{c}}
	{{\bf{P}}_k}{{\bf{\Phi }}({{\bf{R}}_{\bf{S}}})}{{\bf{P}}_k}&{{\bf{P}}_k}\\
	{{\bf{P}}_k}&{{\bf{P}}_k}{{\bf{\Phi }}{{({{\bf{R}}_{\bf{S}}})}^{ - 1}}}{{\bf{P}}_k}
	\end{array}} \right] \succeq 0.
\end{aligned}
\end{equation}
Next, we have
\begin{equation}\small
\begin{aligned}
&\left[ {\begin{array}{*{20}{c}}
	\sum\nolimits_{k=1}^K{{\bf{P}}_k}{{\bf{\Phi }}({{\bf{R}}_{\bf{S}}})}{{\bf{P}}_k}&\sum\nolimits_{k=1}^K{{\bf{P}}_k}\\
	\sum\nolimits_{k=1}^K{{\bf{P}}_k}&\sum\nolimits_{k=1}^K{{\bf{P}}_k}{{\bf{\Phi }}{{({{\bf{R}}_{\bf{S}}})}^{ - 1}}}{{\bf{P}}_k}
	\end{array}} \right]\\
&=\left[ {\begin{array}{*{20}{c}}
	\sum\nolimits_{k=1}^K{{\bf{P}}_k}{{\bf{\Phi }}({{\bf{R}}_{\bf{S}}})}{{\bf{P}}_k}&{\bf I}_{2K}\\
	{\bf I}_{2K}&\sum\nolimits_{k=1}^K{{\bf{P}}_k}{{\bf{\Phi }}{{({{\bf{R}}_{\bf{S}}})}^{ - 1}}}{{\bf{P}}_k}
	\end{array}} \right] \succeq 0.
\end{aligned}
\end{equation}
Then, based on the Schur complement theorem, we can obtain
\begin{equation}\small
\begin{aligned}
&\sum\nolimits_{k=1}^K{{\bf{P}}_k}{{\bf{\Phi }}{{({{\bf{R}}_{\bf{S}}})}^{ - 1}}}{{\bf{P}}_k} - {\bf I}_{2K}\left(\sum\nolimits_{k=1}^K{{\bf{P}}_k}{{\bf{\Phi }}{{({{\bf{R}}_{\bf{S}}})}^{ - 1}}}{{\bf{P}}_k}\right)^{-1}{\bf I}_{2K}\\
&=\sum\limits_{k=1}^K{{\bf{P}}_k}{{\bf{\Phi }}{{({{\bf{R}}_{\bf{S}}})}^{ - 1}}}{{\bf{P}}_k} - \left(\sum\nolimits_{k=1}^K{{\bf{P}}_k}{{\bf{\Phi }}{{({{\bf{R}}_{\bf{S}}})}}}{{\bf{P}}_k}\right)^{-1}\\
&=\sum\nolimits_{k=1}^K{{\bf{P}}_k}{{\bf{\Phi }}{{({{\bf{R}}_{\bf{S}}})}}}{{\bf{P}}_k} - {\bf{\Psi }}({\widetilde{\bf{R}}_{\bf{S}}})^{-1} \succeq 0.
\end{aligned}
\end{equation}
Since $\sum\nolimits_{k=1}^K{{\bf{P}}_k}{{\bf{\Phi }}{{({{\bf{R}}_{\bf{S}}})}^{ - 1}}}{{\bf{P}}_k} - {\bf{\Psi }}({\widetilde{\bf{R}}_{\bf{S}}})$ is a Hermitian matrix and is positive semi-definite, we have
\begin{equation}\label{prove0}\small
\begin{aligned}
&{\rm tr}\left(\sum\nolimits_{k=1}^K{{\bf{P}}_k}{{\bf{\Phi }}{{({{\bf{R}}_{\bf{S}}})}^{ - 1}}}{{\bf{P}}_k} - {\bf{\Psi }}({\widetilde{\bf{R}}_{\bf{S}}})^{-1}\right)\\
&=\sum\nolimits_{k=1}^K{\rm tr}\left({{\bf{P}}_k}{{\bf{\Phi }}{{({{\bf{R}}_{\bf{S}}})}^{ - 1}}}{{\bf{P}}_k}\right) -{\rm tr}\left( {\bf{\Psi }}({\widetilde{\bf{R}}_{\bf{S}}})^{-1}\right)\\
&=\sum\nolimits_{k=1}^K{\rm tr}\left({{\bf{\Phi }}{{({{\bf{R}}_{\bf{S}}})}^{ - 1}}}{{\bf{P}}_k}{{\bf{P}}_k}\right) -{\rm tr}\left( {\bf{\Psi }}({\widetilde{\bf{R}}_{\bf{S}}})^{-1}\right)\\
&={\rm tr}\left({{\bf{\Phi }}{{({{\bf{R}}_{\bf{S}}})}^{ - 1}}}\sum\nolimits_{k=1}^K{{\bf{P}}_k}\right) -{\rm tr}\left( {\bf{\Psi }}({\widetilde{\bf{R}}_{\bf{S}}})^{-1}\right)\\
&={\rm tr}\left({{\bf{\Phi }}{{({{\bf{R}}_{\bf{S}}})}^{ - 1}}}\right) -{\rm tr}\left( {\bf{\Psi }}({\widetilde{\bf{R}}_{\bf{S}}})^{-1}\right) \\
&={\rm tr}\left( {\rm{CRB}}(\tilde{{\bf q}})\right) -{\rm tr}\left( {\bf{\Psi }}({\widetilde{\bf{R}}_{\bf{S}}})^{-1}\right)  \buildrel {\text{(\rm c)}} \over \ge 0,
\end{aligned}
\end{equation}
where the equality at (c) holds if and only if ${\bf{\Phi }}({{\bf{R}}_{\bf{S}}}) ={\bf{\Psi }}({\widetilde{\bf{R}}_{\bf{S}}})$, which is equivalent to (\ref{condition1}). Subsequently, we derive ${\rm tr}\left( {\bf{\Psi }}({\widetilde{\bf{R}}_{\bf{S}}})^{-1}\right)$ in the following.

First, we denote
\begin{equation}\small
{\bf{\Psi }}({\widetilde{\bf{R}}_{\bf{S}}}) = \frac{2}{{{\sigma ^2}}}\left[ {\begin{array}{*{20}{c}}
	{{{\bf{M}}_{11}}}&{{{\bf{M}}_{12}}}\\
	{{{\bf{M}}_{21}}}&{{{\bf{M}}_{22}}}
	\end{array}} \right],
\end{equation}
where
\begin{equation}\small
{{{\bf{M}}_{11}}} = TP_{\rm s}{{\rm{diag}}\left( {{{\left\| {{\bf{\Pi }}_{\bf{A}}^ \bot {{\dot{\bf{ a}}}_{\rm{u}}}\left( {\tilde{\bf{q}},{{\bf{r}}_1}} \right)} \right\|}_2^2}, \cdots ,{{\left\| {{\bf{\Pi }}_{\bf{A}}^ \bot {{\dot{\bf{ a}}}_{\rm{u}}}\left( {\tilde{\bf{q}},{{\bf{r}}_K}} \right)} \right\|}_2^2}} \right)},
\end{equation}
\begin{equation}\small
\begin{aligned}
{{\bf{M}}_{12}} = {{\bf{M}}_{21}} &=TP_{\rm s}
{\rm{diag}}\left( {\Re \left\{ {\dot{\bf{ a}}_{\rm{u}}^{\mathsf H}\left( {\tilde{\bf{q}},{{\bf{r}}_1}} \right){\bf{\Pi }}_{\bf{A}}^ \bot {{\dot{\bf{ a}}}_{\rm{v}}}\left( {{\bf{q}},{{\bf{r}}_1}} \right)} \right\}}, \cdots , \right. \\
&\quad \left. {\Re \left\{ {\dot{\bf{ a}}_{\rm{u}}^{\mathsf H}\left( {\tilde{\bf{q}},{{\bf{r}}_1}} \right){\bf{\Pi }}_{\bf{A}}^ \bot {{\dot{\bf{ a}}}_{\rm{v}}}\left( {{\bf{q}},{{\bf{r}}_K}} \right)} \right\}} \right),
\end{aligned}
\end{equation}
\begin{equation}\small
{{{\bf{M}}_{22}}} = TP_{\rm s}{{\rm{diag}}\left( {{{\left\| {{\bf{\Pi }}_{\bf{A}}^ \bot {{\dot{\bf{ a}}}_{\rm{v}}}\left( {\tilde{\bf{q}},{{\bf{r}}_1}} \right)} \right\|}_2^2}, \cdots ,{{\left\| {{\bf{\Pi }}_{\bf{A}}^ \bot {{\dot{\bf{ a}}}_{\rm{v}}}\left( {\tilde{\bf{q}},{{\bf{r}}_K}} \right)} \right\|}_2^2}} \right)}.
\end{equation}
Then, the diagonal block matrix of $ {\bf{\Psi }}({\widetilde{\bf{R}}_{\bf{S}}})^{-1}$ can be obtained based on the inverse formula of the block matrix as (\ref{inverse1}) and (\ref{inverse2}) shown at the top of the next page.
\begin{figure*}[hbt]
\begin{equation}\label{inverse1} \small
\begin{aligned}
&{({{\bf{M}}_{11}} - {{\bf{M}}_{12}}{{\bf{M}}_{22}^{ - 1}}{{\bf{M}}_{21}})^{ - 1}}=\\
& \frac{1}{TP_{\rm s}}{\rm{diag}}{\left( {{{\left\| {{\bf{\Pi }}_{\bf{A}}^ \bot {{\dot{\bf{ a}}}_{\rm{u}}}\left( {\tilde{\bf{q}},{{\bf{r}}_1}} \right)} \right\|}^2} - \frac{{\Re {{\left\{ {\dot{\bf{ a}}_{\rm{u}}^{\mathsf H}\left( {\tilde{\bf{q}},{{\bf{r}}_1}} \right){\bf{\Pi }}_{\bf{A}}^ \bot {{\dot{\bf{ a}}}_{\rm{v}}}\left( {\tilde{\bf{q}},{{\bf{r}}_1}} \right)} \right\}}^2}}}{{{{\left\| {{\bf{\Pi }}_{\bf{A}}^ \bot {{\dot{\bf{ a}}}_{\rm{v}}}\left( {\tilde{\bf{q}},{{\bf{r}}_1}} \right)} \right\|}^2}}}, \cdots ,{{\left\| {{\bf{\Pi }}_{\bf{A}}^ \bot {{\dot{\bf{ a}}}_{\rm{u}}}\left( {\tilde{\bf{q}},{{\bf{r}}_K}} \right)} \right\|}^2} - \frac{{\Re {{\left\{ {\dot{\bf{ a}}_{\rm{u}}^{\mathsf H}\left( {\tilde{\bf{q}},{{\bf{r}}_K}} \right){\bf{\Pi }}_{\bf{A}}^ \bot {{\dot{\bf{ a}}}_{\rm{v}}}\left( {\tilde{\bf{q}},{{\bf{r}}_K}} \right)} \right\}}^2}}}{{{{\left\| {{\bf{\Pi }}_{\bf{A}}^ \bot {{\dot{\bf{ a}}}_{\rm{v}}}\left( {\tilde{\bf{q}},{{\bf{r}}_K}} \right)} \right\|}^2}}}} \right)^{ - 1}}
\end{aligned}
\end{equation}
\vspace*{-1em} 
\hrulefill
\end{figure*}
\begin{figure*}[hbt]
	\begin{equation}\label{inverse2}\small
	\begin{aligned}
	&{({{\bf{M}}_{22}} - {{\bf{M}}_{21}}{{\bf{M}}_{11}^{ - 1}}{{\bf{M}}_{12}})^{ - 1}}=\\
	& \frac{1}{TP_{\rm s}}{\rm{diag}}{\left( {{{\left\| {{\bf{\Pi }}_{\bf{A}}^ \bot {{\dot{\bf{ a}}}_{\rm{v}}}\left( {\tilde{\bf{q}},{{\bf{r}}_1}} \right)} \right\|}^2} - \frac{{\Re {{\left\{ {\dot{\bf{ a}}_{\rm{u}}^{\mathsf H}\left( {\tilde{\bf{q}},{{\bf{r}}_1}} \right){\bf{\Pi }}_{\bf{A}}^ \bot {{\dot{\bf{ a}}}_{\rm{v}}}\left( {\tilde{\bf{q}},{{\bf{r}}_1}} \right)} \right\}}^2}}}{{{{\left\| {{\bf{\Pi }}_{\bf{A}}^ \bot {{\dot{\bf{ a}}}_{\rm{u}}}\left( {\tilde{\bf{q}},{{\bf{r}}_1}} \right)} \right\|}^2}}}, \cdots ,{{\left\| {{\bf{\Pi }}_{\bf{A}}^ \bot {{\dot{\bf{ a}}}_{\rm{v}}}\left( {\tilde{\bf{q}},{{\bf{r}}_K}} \right)} \right\|}^2} - \frac{{\Re {{\left\{ {\dot{\bf{ a}}_{\rm{u}}^{\mathsf H}\left( {\tilde{\bf{q}},{{\bf{r}}_K}} \right){\bf{\Pi }}_{\bf{A}}^ \bot {{\dot{\bf{ a}}}_{\rm{v}}}\left( {\tilde{\bf{q}},{{\bf{r}}_K}} \right)} \right\}}^2}}}{{{{\left\| {{\bf{\Pi }}_{\bf{A}}^ \bot {{\dot{\bf{ a}}}_{\rm{u}}}\left( {\tilde{\bf{q}},{{\bf{r}}_K}} \right)} \right\|}^2}}}} \right)^{ - 1}}
	\end{aligned}
	\end{equation}
	\vspace*{-1em} 
	\hrulefill
\end{figure*}
As such,  ${\rm tr}\left( {\bf{\Psi }}({\widetilde{\bf{R}}_{\bf{S}}})^{-1}\right)$ can be given by
\begin{equation}\label{prove}\small
\begin{aligned}
&{\rm tr}\left( {\bf{\Psi }}({\widetilde{\bf{R}}_{\bf{S}}})^{-1}\right) = \frac{{{\sigma ^2}}}{2} {\rm tr} \left({\left({{\bf{M}}_{11}} - {{\bf{M}}_{12}}{{\bf{M}}_{22}^{ - 1}}{{\bf{M}}_{21}}\right)^{ - 1}} \right) \\
&+ \frac{{{\sigma ^2}}}{2}{\rm tr}\left({\left({{\bf{M}}_{22}} - {{\bf{M}}_{21}}{{\bf{M}}_{11}^{ - 1}}{{\bf{M}}_{12}}\right)^{ - 1}}\right)\\
&= \frac{{{\sigma ^2}}}{2TP_{\rm s}} \sum\limits_{k=1}^{K}\frac{1}{{{\left\| {{\bf{\Pi }}_{\bf{A}}^ \bot {{\dot{\bf{ a}}}_{\rm{u}}}\left( {\tilde{\bf{q}},{{\bf{r}}_k}} \right)} \right\|}^2} - \frac{{\Re {{\left\{ {\dot{\bf{ a}}_{\rm{u}}^{\mathsf H}\left( {\tilde{\bf{q}},{{\bf{r}}_k}} \right){\bf{\Pi }}_{\bf{A}}^ \bot {{\dot{\bf{ a}}}_{\rm{v}}}\left( {\tilde{\bf{q}},{{\bf{r}}_k}} \right)} \right\}}^2}}}{{{{\left\| {{\bf{\Pi }}_{\bf{A}}^ \bot {{\dot{\bf{ a}}}_{\rm{v}}}\left( {\tilde{\bf{q}},{{\bf{r}}_k}} \right)} \right\|}^2}}}}\\
&+\frac{{{\sigma ^2}}}{2TP_{\rm s}} \sum\limits_{k=1}^{K}\frac{1}{{{\left\| {{\bf{\Pi }}_{\bf{A}}^ \bot {{\dot{\bf{ a}}}_{\rm{v}}}\left( {\tilde{\bf{q}},{{\bf{r}}_k}} \right)} \right\|}^2} - \frac{{\Re {{\left\{ {\dot{\bf{ a}}_{\rm{u}}^{\mathsf H}\left( {\tilde{\bf{q}},{{\bf{r}}_k}} \right){\bf{\Pi }}_{\bf{A}}^ \bot {{\dot{\bf{ a}}}_{\rm{v}}}\left( {\tilde{\bf{q}},{{\bf{r}}_k}} \right)} \right\}}^2}}}{{{{\left\| {{\bf{\Pi }}_{\bf{A}}^ \bot {{\dot{\bf{ a}}}_{\rm{u}}}\left( {\tilde{\bf{q}},{{\bf{r}}_k}} \right)} \right\|}^2}}}}.
\end{aligned}
\end{equation}

Next, we rewrite
\begin{equation}\label{begin}\small
\begin{aligned}
&{{\left\| {{\bf{\Pi }}_{\bf{A}}^ \bot {{\dot{\bf{ a}}}_{\rm{u}}}\left( {\tilde{\bf{q}},{{\bf{r}}_k}} \right)} \right\|}^2} - \frac{{\Re {{\left\{ {\dot{\bf{ a}}_{\rm{u}}^{\mathsf H}\left( {\tilde{\bf{q}},{{\bf{r}}_k}} \right){\bf{\Pi }}_{\bf{A}}^ \bot {{\dot{\bf{ a}}}_{\rm{v}}}\left( {\tilde{\bf{q}},{{\bf{r}}_k}} \right)} \right\}}^2}}}{{{{\left\| {{\bf{\Pi }}_{\bf{A}}^ \bot {{\dot{\bf{ a}}}_{\rm{v}}}\left( {\tilde{\bf{q}},{{\bf{r}}_k}} \right)} \right\|}^2}}}\\
&=  \mathop{\min }\limits_{\zeta_{{\rm u},k} \in \mathbb R} {\left\| {{\bf{\Pi }}_{\bf{A}}^ \bot \left( {{{\dot{\bf{ a}}}_{\rm{u}}}\left( {\tilde{\bf{q}},{{\bf{r}}_k}} \right) - \zeta_{{\rm u},k} {{\dot{\bf{ a}}}_{\rm{v}}}\left( {\tilde{\bf{q}},{{\bf{r}}_k}} \right)} \right)} \right\|}_2^2,
\end{aligned}
\end{equation}
which can be easily proved via convex quadratic optimization and is omitted here. Then, we denote the subspace $\bf U$ spanned by vector ${{\bf{ a}}}\left( {\tilde{\bf{q}},{{\bf{r}}_k}}\right)$ as ${\bf U} = {\rm span} ({{\bf{ a}}}\left( {\tilde{\bf{q}},{{\bf{r}}_k}}\right))$ and the subspace $\bf V$ spanned by the column vectors of $\bf A$ as ${\bf V} = {\rm col}(\bf A)$, with their projection matrix respectively given by ${\bf \Pi}_{k}$ and ${\bf \Pi}_{\bf A}$. Since the basis vectors of $\mathbf{U}$ are in $\mathbf{V}$, $\bf U$ is a subspace of $\bf V$, i.e., ${\bf U} \subseteq {\bf V}$. Denote ${\bf W} = {\bf U}^{^ \bot} \cap {\bf V}$ with ${\bf U}^{^ \bot}$ representing the orthogonal subspace to $\bf U$ and denote its projection matrix as ${\bf \Pi}_{\bf W}$. Then, we have
\begin{equation}\small
{\bf \Pi}_{\bf A}^{\bot} = {\bf I}_N - {\bf \Pi}_{\bf A} = ({\bf I}_N - {\bf \Pi}_{\bf W})({\bf I}_N - {\bf \Pi}_{k}),
\end{equation}
since ${\bf \Pi}_{\bf W}{\bf \Pi}_{k} = {\bf 0}_N$. Then, for $\zeta_{{\rm u},k} \in \mathbb R$, we have (\ref{prove1}) shown at the top of this page, where the equality at (d) holds if and only if
\begin{figure*}[t]
\begin{equation} \label{prove1} \small
\begin{aligned}
&\mathop{\min }\limits_{\zeta_{{\rm u},k} \in \mathbb R} {\left\| {{\bf{\Pi }}_{\bf{A}}^ \bot \left( {{{\dot{\bf{ a}}}_{\rm{u}}}\left( {\tilde{\bf{q}},{{\bf{r}}_k}} \right) - \zeta_{{\rm u},k} {{\dot{\bf{ a}}}_{\rm{v}}}\left( {\tilde{\bf{q}},{{\bf{r}}_k}} \right)} \right)} \right\|}_2^2\\
&\buildrel {\text{(\rm d)}} \over \le \mathop{\min }\limits_{\zeta_{{\rm u},k} \in \mathbb R} {\left\| {({\bf I}_N - {\bf \Pi}_{k}) \left( {{{\dot{\bf{ a}}}_{\rm{u}}}\left( {\tilde{\bf{q}},{{\bf{r}}_k}} \right) - \zeta_{{\rm u},k} {{\dot{\bf{ a}}}_{\rm{v}}}\left( {\tilde{\bf{q}},{{\bf{r}}_k}} \right)} \right)} \right\|}_2^2 = {\left\| {{\bf{\Pi }}_k^ \bot {{\dot{\bf{ a}}}_{\rm{u}}}\left( {\tilde{\bf{q}},{{\bf{r}}_k}} \right)} \right\|^2} - \frac{{\Re {{\left\{ {\dot{\bf{ a}}_{\rm{u}}^{\mathsf H}\left( {\tilde{\bf{q}},{{\bf{r}}_k}} \right){\bf{\Pi }}_k^ \bot {{\dot{\bf{ a}}}_{\rm{v}}}\left( {\tilde{\bf{q}},{{\bf{r}}_k}} \right)} \right\}}^2}}}{{{{\left\| {{\bf{\Pi }}_k^ \bot {{\dot{\bf{ a}}}_{\rm{v}}}\left( {\tilde{\bf{q}},{{\bf{r}}_k}} \right)} \right\|}^2}}}\\
&= \dot{\bf{ a}}_{\rm{u}}^{\mathsf H}\left( {\tilde{\bf{q}},{{\bf{r}}_k}} \right)\left( {{\bf{I}}_N - \frac{{{{\bf{a}}\left( {\tilde{\bf{q}},{{\bf{r}}_k}} \right)}{\bf{a}}\left( {\tilde{\bf{q}},{{\bf{r}}_k}} \right)^{\mathsf H}}}{{{{\left\| {{{\bf{a}}\left( {\tilde{\bf{q}},{{\bf{r}}_k}} \right)}} \right\|}^2}}}} \right){{\dot{\bf{ a}}}_{\rm{u}}}\left( {\tilde{\bf{q}},{{\bf{r}}_k}} \right)  - \frac{{{{\left( {\dot{\bf{ a}}_{\rm{u}}^{\mathsf H}\left( {\tilde{\bf{q}},{{\bf{r}}_k}} \right)\left( {{\bf{I}}_N - \frac{{{{\bf{a}}\left( {\tilde{\bf{q}},{{\bf{r}}_k}} \right)}{\bf{a}}\left( {\tilde{\bf{q}},{{\bf{r}}_k}} \right)^{\mathsf H}}}{{{{\left\| {{{\bf{a}}\left( {\tilde{\bf{q}},{{\bf{r}}_k}} \right)}} \right\|}^2}}}} \right){{\dot{\bf{ a}}}_{\rm{v}}}\left( {\tilde{\bf{q}},{{\bf{r}}_k}} \right)} \right)}^2}}}{{\dot{\bf{ a}}_{\rm{v}}^{\mathsf H}\left( {\tilde{\bf{q}},{{\bf{r}}_k}} \right)\left( {{\bf{I}}_N - \frac{{{{\bf{a}}\left( {\tilde{\bf{q}},{{\bf{r}}_k}} \right)}{\bf{a}}\left( {\tilde{\bf{q}},{{\bf{r}}_k}} \right)^{\mathsf H}}}{{{{\left\| {{{\bf{a}}\left( {\tilde{\bf{q}},{{\bf{r}}_k}} \right)}} \right\|}^2}}}} \right){{\dot{\bf{ a}}}_{\rm{v}}}\left( {\tilde{\bf{q}},{{\bf{r}}_k}} \right)}}\\
& = \frac{{4{\pi ^2}}}{{{\lambda ^2}}}\left( \sum\limits_{n = 1}^N {x_n^2 - \frac{1}{N}} {\left( {\sum\limits_{n = 1}^N {{x_n}} } \right)^2} - \frac{{{{\left( {\sum\limits_{n = 1}^N {{x_n}{y_n}}  - \frac{1}{N}\sum\limits_{n = 1}^N {{x_n}} \sum\limits_{n = 1}^N {{y_n}} } \right)}^2}}}{{\sum\limits_{n = 1}^N {y_n^2 - \frac{1}{N}} {{\left( {\sum\limits_{n = 1}^N {{y_n}} } \right)}^2}}}\right) = \frac{{4N{\pi ^2}}}{{{\lambda ^2}}} \left({{{\rm{var}}(\tilde{\bf{x}}) - \frac{{{{\rm cov}} {{(\tilde{ \bf{x}},\tilde{\bf{y}})}^2}}}{{{\rm{var}}(\tilde{\bf{y}})}}}}\right)
\end{aligned}
\end{equation}
	\vspace*{-1em} 
\hrulefill
\end{figure*}
\begin{equation} \label{nece1}
{{\bf{\Pi}}_{\bf{W}}}{\bf{\Pi }}_k^ \bot \left( {{{\dot{\bf{ a}}}_{\rm{u}}}\left( {\tilde{\bf{q}},{{\bf{r}}_k}} \right) - {\zeta_{{\rm u},k} ^*}{{\dot{\bf{ a}}}_{\rm{v}}}\left( {\tilde{\bf{q}},{{\bf{r}}_k}} \right)} \right) = {\bf{0}}_{N \times 1}.
\end{equation}
Moreover, given that ${\bf U} \subseteq {\bf V}$ and ${\bf V}$ is the orthogonal direct sum of $\bf U$ and $\bf W$, i.e., ${\bf V} = {\bf U} \oplus {\bf W}$, we have
\begin{equation}\small
{\bf{\Pi}}_{\bf{A}} = {\bf{\Pi}}_{k} +{\bf{\Pi}}_{\bf{W}}.
\end{equation}
Furthermore, since ${\bf{\Pi}}_{k}{\bf{\Pi}}_{k}^{\bot} = {\bf 0}_N$, we have
\begin{equation}\small
{\bf{\Pi}}_{\bf{A}}{\bf{\Pi}}_{k}^{\bot} = {\bf{\Pi}}_{\bf{W}}{\bf{\Pi}}_{k}^{\bot}.
\end{equation}
Next, we have
\begin{equation}\small
\begin{aligned}
&{\bf{\Pi}}_{\bf{A}}{\bf{\Pi}}_{k}^{\bot}\left( {{{\dot{\bf{ a}}}_{\rm{u}}}\left( {\tilde{\bf{q}},{{\bf{r}}_k}} \right) - {\zeta_{{\rm u},k} ^*}{{\dot{\bf{ a}}}_{\rm{v}}}\left( {\tilde{\bf{q}},{{\bf{r}}_k}} \right)} \right)\\
&= {{\bf{\Pi}}_{\bf{W}}}{\bf{\Pi }}_k^ \bot \left( {{{\dot{\bf{ a}}}_{\rm{u}}}\left( {\tilde{\bf{q}},{{\bf{r}}_k}} \right) - {\zeta_{{\rm u},k} ^*}{{\dot{\bf{ a}}}_{\rm{v}}}\left( {\tilde{\bf{q}},{{\bf{r}}_k}} \right)} \right) = {\bf{0}}_{N \times 1}.
\end{aligned}
\end{equation}
In addition, since the null space of $\bf A$ is equivalent to that of ${\bf{\Pi}}_{\bf{A}}$, i.e., ${\rm range}({\bf{A}}) = {\rm range}({\bf{\Pi}}_{\bf{A}})$, we have
\begin{equation}\label{nece2}
\begin{aligned}
&{\bf{A}}^{\mathsf H}{\bf{\Pi}}_{k}^{\bot}\left( {{{\dot{\bf{ a}}}_{\rm{u}}}\left( {\tilde{\bf{q}},{{\bf{r}}_k}} \right) - {\zeta_{{\rm u},k} ^*}{{\dot{\bf{ a}}}_{\rm{v}}}\left( {\tilde{\bf{q}},{{\bf{r}}_k}} \right)} \right) = {\bf{0}}_{K \times 1}.
\end{aligned}
\end{equation}
In other words, the sufficient and necessary condition for the equality at (d) in (\ref{nece1}) is equivalent to (\ref{nece2}).

Similar to the procedure from (\ref{begin}) to (\ref{nece2}), we have
\begin{equation}\label{prove2}\small
\begin{aligned}
&{{\left\| {{\bf{\Pi }}_{\bf{A}}^ \bot {{\dot{\bf{ a}}}_{\rm{v}}}\left( {\tilde{\bf{q}},{{\bf{r}}_k}} \right)} \right\|}^2} - \frac{{\Re {{\left\{ {\dot{\bf{ a}}_{\rm{u}}^{\mathsf H}\left( {\tilde{\bf{q}},{{\bf{r}}_k}} \right){\bf{\Pi }}_{\bf{A}}^ \bot {{\dot{\bf{ a}}}_{\rm{v}}}\left( {\tilde{\bf{q}},{{\bf{r}}_k}} \right)} \right\}}^2}}}{{{{\left\| {{\bf{\Pi }}_{\bf{A}}^ \bot {{\dot{\bf{ a}}}_{\rm{u}}}\left( {\tilde{\bf{q}},{{\bf{r}}_k}} \right)} \right\|}^2}}}\\
&=  \mathop{\min }\limits_{\zeta_{{\rm v},k} \in \mathbb R} {\left\| {{\bf{\Pi }}_{\bf{A}}^ \bot \left( {{{\dot{\bf{ a}}}_{\rm{v}}}\left( {\tilde{\bf{q}},{{\bf{r}}_k}} \right) - \zeta_{{\rm v},k} {{\dot{\bf{ a}}}_{\rm{u}}}\left( {\tilde{\bf{q}},{{\bf{r}}_k}} \right)} \right)} \right\|}_2^2\\
&\buildrel {\text{(\rm e)}} \over \le \frac{{4N{\pi ^2}}}{{{\lambda ^2}}} \left({{{\rm{var}}(\tilde{\bf{y}}) - \frac{{{{\rm cov}} {{(\tilde{ \bf{x}},\tilde{\bf{y}})}^2}}}{{{\rm{var}}(\tilde{\bf{x}})}}}}\right),
\end{aligned}
\end{equation}
where the equality at (e) holds if and only if
\begin{equation}\label{nece3}\small
\begin{aligned}
&{\bf{A}}^{\mathsf H}{\bf{\Pi}}_{k}^{\bot}\left( {{{\dot{\bf{ a}}}_{\rm{v}}}\left( {\tilde{\bf{q}},{{\bf{r}}_k}} \right) - {\zeta_{{\rm v},k} ^*}{{\dot{\bf{ a}}}_{\rm{u}}}\left( {\tilde{\bf{q}},{{\bf{r}}_k}} \right)} \right) = {\bf{0}}_{K \times 1}.
\end{aligned}
\end{equation}
Hereby, by combining (\ref{prove0}), (\ref{prove}), (\ref{prove1}), and (\ref{prove2}), we have the first inequality in (\ref{lower}), i.e., (a), where the equality holds if and only if the equalities at (c) , (d), and (e) hold simultaneously.

Next, we prove that (b) in (\ref{lower}) holds. Specifically, we have
\begin{equation}\small
\begin{aligned}
 & \quad {\frac{1}{{{\rm{var}}(\tilde{\bf{x}}) - \frac{{{{\rm cov}} {{(\tilde{ \bf{x}},\tilde{\bf{y}})}^2}}}{{{\rm{var}}(\tilde{\bf{y}})}}}} + \frac{1}{{{\rm{var}}(\tilde{\bf{y}}) - \frac{{{{\rm cov}} {{(\tilde{\bf{x}},\tilde{\bf{y}})}^2}}}{{{\rm{var}}(\tilde{\bf{x}})}}}}}\\
 &\ge \frac{4}{{{{\rm{var}}(\tilde{\bf{x}}) + {\rm{var}}(\tilde{\bf{y}}) - \frac{{{{\rm cov}} {{(\tilde{ \bf{x}},\tilde{\bf{y}})}^2}}}{{{\rm{var}}(\tilde{\bf{x}})}}}} - \frac{{{{\rm cov}} {{(\tilde{ \bf{x}},\tilde{\bf{y}})}^2}}}{{{\rm{var}}(\tilde{\bf{y}})}}}\\
 &\ge \frac{4}{{\rm{var}}(\tilde{\bf{x}}) + {\rm{var}}(\tilde{\bf{y}})} = \frac{4}{\frac{1}{N}\sum\limits_{n=1}^{N}x_n^2 - \mu(\tilde{\bf{x}})^2 + \frac{1}{N}\sum\limits_{n=1}^{N}y_n^2 - \mu(\tilde{\bf{y}})^2}\\
 &\ge \frac{4N}{\sum\limits_{n=1}^{N}(x_n^2 + y_n^2)} \buildrel {\text{(\rm f)}} \over \ge \frac{8}{A^2},
\end{aligned}
\end{equation}
where the equality at (f) holds if and only is (\ref{condition3}) holds.

This completes the proof.
\linespread{1}
\bibliographystyle{IEEEtran} 
\bibliography{MA_for_Multi_Target_Sensing}

\vfill

\end{document}